\documentclass[12pt]{article}
 
\usepackage{latexsym}
\usepackage{epsfig}
\usepackage{psfig}

\begin{document}
\begin{center}
{\Large\bf
QCD in a finite box: Numerical test studies\\
\vspace*{3pt}
in the three Leutwyler-Smilga regimes}
\end{center}
\vspace*{6pt}

\begin{center}
{\bf Stephan D\"urr}
\vspace*{3pt}\\
{\sl Paul Scherrer Institut, Theory Group}\\
{\sl 5232 Villigen PSI, Switzerland}\\
\vspace*{1pt}
{\tt stephan.duerr@psi.ch}
\end{center}
\vspace*{8pt}

\begin{abstract}
The Leutwyler-Smilga prediction regarding the (ir)relevance of the global
topological charge for QCD in a finite box is subject to a test. To this end
the lattice version of a suitably chosen analogue (massive 2-flavour Schwinger
model) is analyzed in the small ($V\Sigma m\ll 1$), intermediate ($V\Sigma m
\simeq 1$) and large ($V\Sigma m\gg 1$) Leutwyler-Smilga regimes. The
predictions for the small and large regimes are confirmed and illustrated. New
results about the role of the functional determinant in all three regimes and
about the sensitivity of physical observables on the topological charge in the
intermediate regime are presented.
\end{abstract}
\vspace*{6pt}


\newcommand{\pad}{\partial}
\newcommand{\pas}{\partial\!\!\!/}
\newcommand{\Dsl}{D\!\!\!\!/\,}
\newcommand{\Psl}{P\!\!\!\!/\;\!}
\newcommand{\hqu}{\hbar}
\newcommand{\ovr}{\over} 
\newcommand{\til}{\tilde}
\newcommand{\pri}{^\prime}
\renewcommand{\dag}{^\dagger}
\newcommand{\<}{\langle}
\renewcommand{\>}{\rangle}
\newcommand{\gaf}{\gamma_5}
\newcommand{\lap}{\triangle}
\newcommand{\trc}{\rm tr}
\newcommand{\al}{\alpha}
\newcommand{\be}{\beta}
\newcommand{\ga}{\gamma}
\newcommand{\de}{\delta}
\newcommand{\ep}{\epsilon}
\newcommand{\ve}{\varepsilon}
\newcommand{\ze}{\zeta}
\newcommand{\et}{\eta}
\renewcommand{\th}{\theta}
\newcommand{\vt}{\vartheta}
\newcommand{\io}{\iota}
\newcommand{\ka}{\kappa}
\newcommand{\la}{\lambda}
\newcommand{\rh}{\rho}
\newcommand{\vr}{\varrho}
\newcommand{\si}{\sigma}
\newcommand{\ta}{\tau}
\newcommand{\ph}{\phi}
\newcommand{\vp}{\varphi}
\newcommand{\ch}{\chi}
\newcommand{\ps}{\psi}
\newcommand{\om}{\omega}
\newcommand{\psb}{\overline{\psi}}
\newcommand{\etb}{\overline{\eta}}
\newcommand{\psd}{\psi^{\dagger}}
\newcommand{\etd}{\eta^{\dagger}}
\newcommand{\beq}{\begin{equation}}
\newcommand{\eeq}{\end{equation}}
\newcommand{\bdm}{\begin{displaymath}}
\newcommand{\edm}{\end{displaymath}}
\newcommand{\bea}{\begin{eqnarray}}
\newcommand{\eea}{\end{eqnarray}}


\section{Introduction}

One of the relevant concepts in an attempt to understand the mechanism of
spontaneous breakdown of chiral symmetry%
\footnote{We shall only consider the theory with several light dynamical
fermions ($N_{\!f}\geq2$).}
in QCD is provided by instantons, i.e.\ topologically nontrivial solutions of
the classical field equations which are both a local minimum of the classical
action and localized in space-time.
They are known to influence the {\em local\/} propagation properties of the
light flavours (Instanton Liquid Model \cite{SchaferShuryak}).

What is not so clear is whether the {\em global\/} concept of the number of
instantons minus anti-instantons for QCD in a finite box
\beq
\nu={g^2\ovr 32\pi^2}\int G\til G\ dx
\label{nudef}
\eeq
plays a role, too.
This is important, because in lattice studies involving dynamical fermions
standard simulation algorithms tend to get ``stuck'' in a particular
topological sector if the sea-quarks are taken sufficiently light%
\footnote{The phenomenon was first observed for HMC with staggered quarks
\cite{TopErgodicityHMCstag}. For Wilson type sea-quarks mobility between the
topological sectors was reported to be sufficient for $M_\pi/M_\rh\!>\!0.56$,
but the same problem emerges if $\kappa$ is tuned sufficiently close to
$\kappa_{\rm crit}$ \cite{TopErgodicityHMCwils, AllesICHEP96}.}
and obviously one would like to know whether this affects physical observables.
In other words, the question is whether non-ergodicities of the sample w.r.t.\
the topological charge as visible e.g.\ from the r.h.s.\ of Fig.~1 tend to
afflict measurements performed on such a sample.

\begin{figure}
\vspace*{-10mm}
\hspace*{-5mm}
\includegraphics{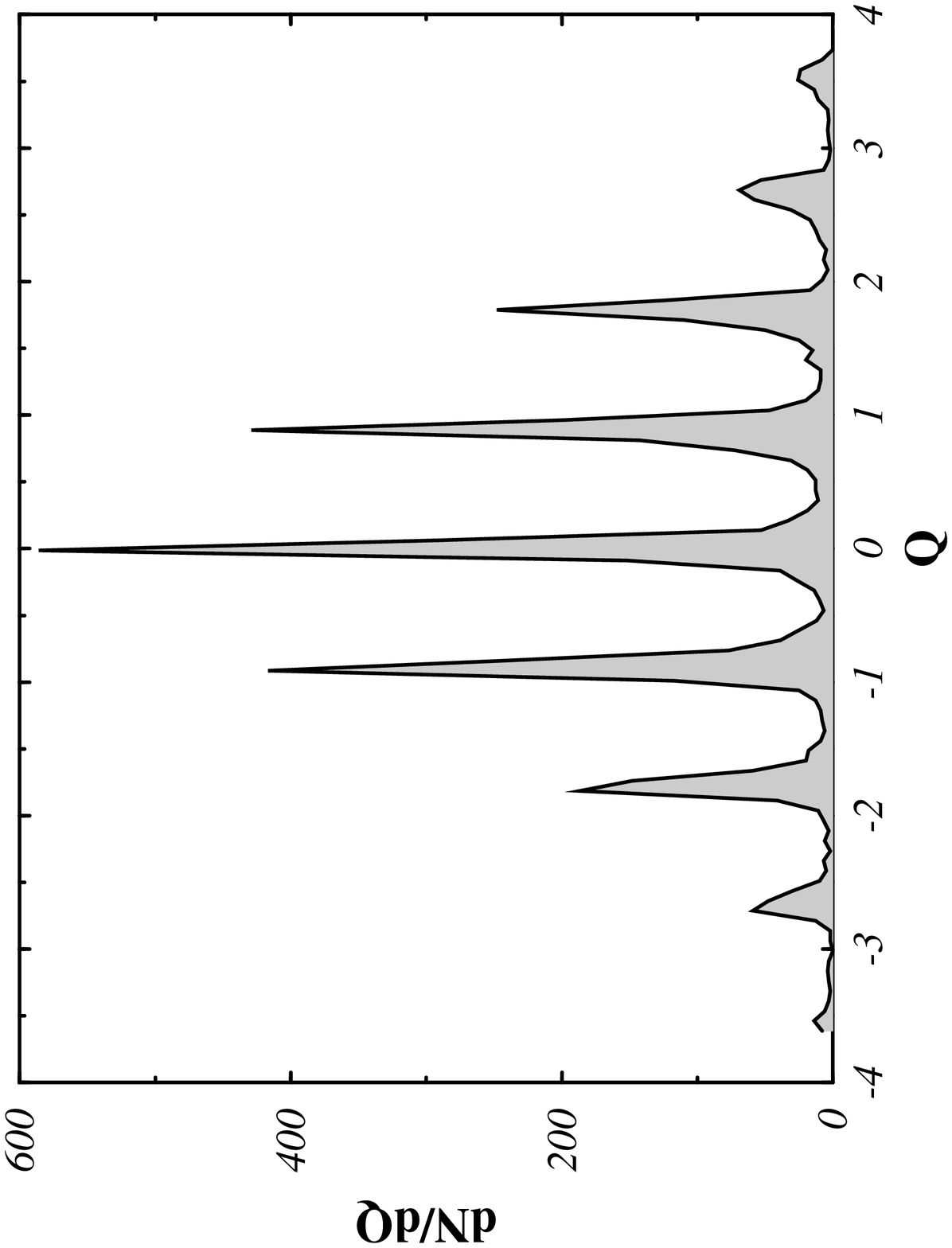}
\includegraphics{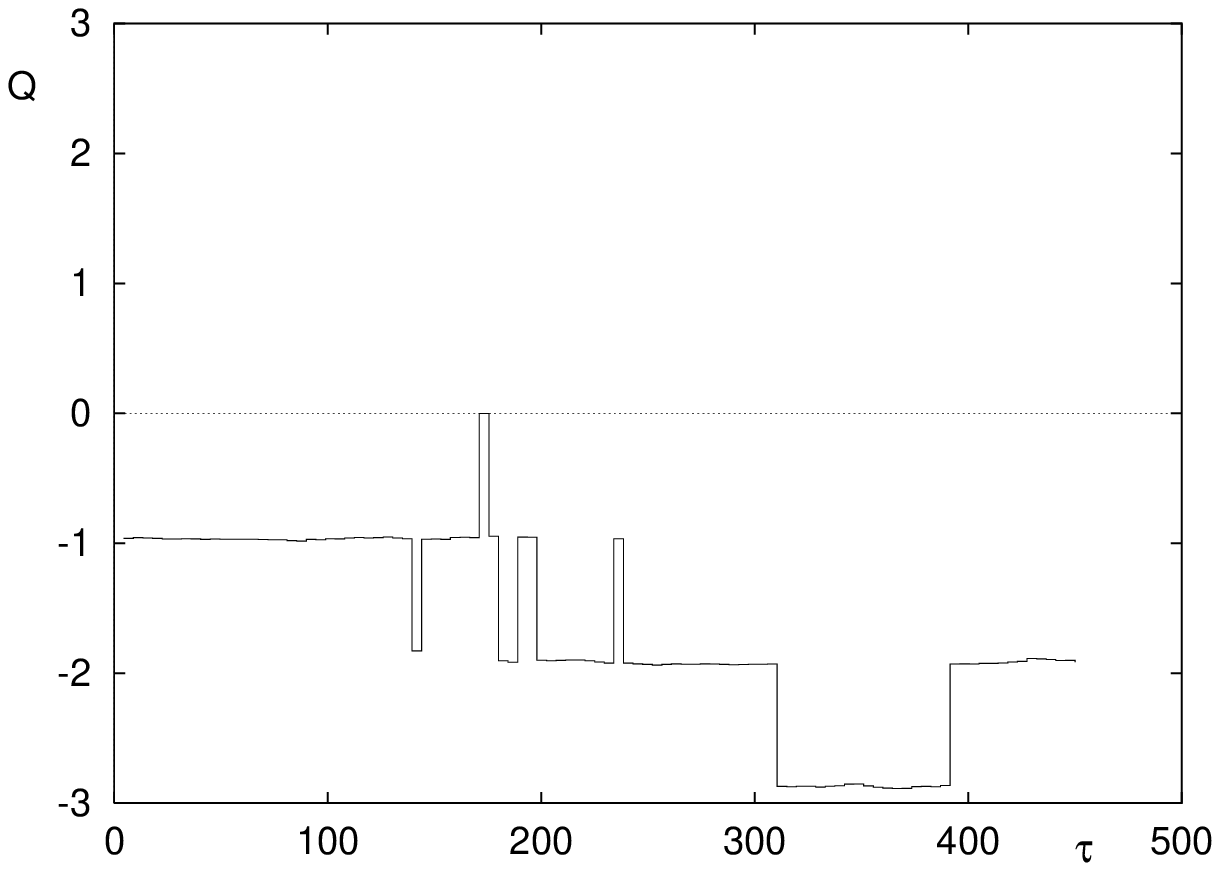}
\vspace*{52mm}
\caption{\sl\small
LHS: Distribution of the unrenormalized field-theoretic topological charge
$\nu_{\rm nai}$ in an ensemble of 5000 $SU(3)$-configurations generated from
$S_{\rm Wilson}$ at $\beta\!=\!6.1$, after 6 cooling sweeps and with a
2-smeared charge operator (figure taken from \cite{AllesICHEP96}).
RHS: Time history, in units of MD time, of $\nu_{\rm nai}$ for a HMC simulation
with a quadruple of staggered quarks at $\beta\!=\!5.35, m\!=\!0.01$; the
analogous distribution lacks the symmetry $\nu\!\leftrightarrow\!-\nu$ and
shows $\<\nu\>\!\neq\!0$ (figure taken from \cite{AllesICHEP96}).}
\end{figure}

The correct sampling of the topological sectors is known to be crucial for
quantities such as the $\et'$ mass, which depend directly on the distribution
of topological charges via the explicit breaking of the $U(1)_A$ symmetry
\cite{LatticeU1Astudies}.
On the other hand, standard observables which do not relate to the $U(1)_A$
issue ($M_\pi,M_K,M_\eta,M_\rho,V_{q\bar q},\ldots$) are known not to depend on
the global topological charge $\nu$ if the four-volume $V$ of the box is taken
sufficiently large.
An early investigation trying to assess the issue on a more quantitative level
is the one by Leutwyler and Smilga \cite{LeutwylerSmilga}.
Their main observation is that for the case of pionic observables a complete
answer can be given from purely analytic considerations for two extreme regimes
of quark masses and box volumes, both of which involve the Leutwyler-Smilga
(LS) parameter%
\footnote{$V$ is the four-volume of the box,
$\Sigma\!=\!-\!\lim_{m\to 0}\lim_{V\to\infty} \langle\bar\psi\psi\rangle$
(this order) is the chiral condensate in the chiral limit and $m$ is the
(degenerate) {\em sea\/}-quark mass; note that both $\Sigma$ and $m$ are
scheme- and scale-dependent, but the combination is an RG-invariant quantity.}
\beq
x\equiv V\Sigma m
\;.
\label{LSPdef}
\eeq
Their first statement is that in a sufficiently small box ($x\ll1$) the
question whether the algorithm managed to tunnel between the various
topological sectors at a sufficient rate shouldn't emerge, because the
partition function is completely dominated by the topologically trivial sector.
In lattice language this amounts to the statement that the algorithm is not
supposed to tunnel into a nontrivial sector anyways.
Their second statement is that in the opposite regime ($x\gg1$ plus a condition
to be discussed below) the total topological charge is an irrelevant concept.
This means, in lattice language, that the observable is unaffected by
whichever ``topological path'' the simulation followed.
From a lattice perspective, however, the analysis by Leutwyler and Smilga is
not quite sufficient, as their paper doesn't state explicitly whether a
simulation in the small $x$ regime can be trusted, if the algorithm got stuck
in a higher topological sector even though it was not supposed to get into it,
and also because there is no statement about the situation in the intermediate
($x\simeq1$) regime, which might be interesting to use in dynamical
simulations.
Last but not least the relevance of the second condition in the LS-analysis for
large $x$ (a condition which, as we shall see, is never fulfilled in QCD
simulations) is not discussed.

Below numerical results will be used to elucidate in which regimes of box
volumes and quark masses a standard observable like the heavy quark potential
is indeed insensitive to the global topological charge and by which mechanism a
dependence sneaks in if the ``insensitive'' regime is left.
I start with a brief review of the approach taken by Leutwyler and Smilga,
followed by an argument why the massive multi-flavour Schwinger model is a
suitable testbed for numerical studies.
Sections 4 through 6 present distinctive features by which the three LS-regimes
differ from each other~--- both on a formal level (role being played by the
functional determinant) and w.r.t.\ observables (Polyakov loop, heavy quark
potential).
The strategy used is to disentangle the complete sample into contributions from
the individual topological sectors and to observe how the generating functional
weights the fixed-$\nu$ expectation values to form the physical
observable~\cite{Damgaard}.
I conclude with a few remarks on the potential relevance of the results for
future QCD simulations.


\section{Leutwyler-Smilga analysis for QCD}

Here I shall briefly summarize the considerations by Leutwyler and Smilga
regarding the distribution of topological charges for QCD in a finite box
\cite{LeutwylerSmilga}.

Leutwyler and Smilga start from the partition function for QCD with
$N_{\!f}\!\geq\!2$ degenerate light (dynamical) flavours on an euclidean torus
\beq
Z\sim\sum_{\nu\in\mathbf{Z}}\int DA^{(\nu)}\;
\det(\Dsl\!+\!m)^{N_{\!f}}\;e^{-(1/4)\int GG}
\label{PFfund}
\eeq
where the integration is over all gauge potentials in a given topological
sector and the quark fields have been integrated out.
In (\ref{PFfund}) the Vafa Witten representation \cite{VafaWitten}
for the functional determinant
\beq
\det(\Dsl\!+\!m)^{N_{\!f}}=
\left(m^{|\nu|}\;{\prod}'(\la^2\!+\!m^2)\right)^{N_{\!f}}
\label{VafaWittenFormula}
\eeq
may be used which splits the latter into a factor representing the contribution
from the zero modes of the Dirac operator $\Dsl$ on the given background times
a chain of factors $\pm\mathrm{i}\la\!+\!m$ from the paired non-zero modes.
Upon using the nontrivial input for the first few non-zero eigenvalues
\beq
\la_n\simeq{n\pi\over V\Sigma}
\label{BanksCasherFormula}
\eeq
(which is just the familiar Banks Casher relation \cite{BanksCasher}) the
primed factor in the decomposition (\ref{VafaWittenFormula}) is seen not to
depend on $m$ if $x\!\ll\!1$, and hence
\beq
Z_\nu\sim m^{|\nu|N_{\!f}}
\qquad\qquad(x\ll1)
\;.
\label{LSsmall}
\eeq
This means that in the chiral symmetry restoration regime ($x\!\ll\!1$)
topologically nontrivial sectors are heavily suppressed w.r.t.\ the trivial
one~--- in particular for small quark masses (in lattice units) and large
$N_{\!f}$.

On the other hand, if the box volume is sufficiently large, the onset of
chiral symmetry breaking gets manifest as the long-range Green's functions get
dominated by the Goldstone excitations and the theory is effectively described
by a chiral lagrangian%
\footnote{References to all the subtle points (e.g.\ why in the particular case
of the torus no boundary terms show up in $L_\mathrm{XPT}$) are found in
\cite{LeutwylerSmilga}.}
\beq
Z\sim\int_{SU(N_{\!f})}DU\;e^{-\int L_\mathrm{XPT}}
\;,
\quad
L_\mathrm{XPT}={F^2\ovr4}\<\pad U \pad U\dag\>-{\Sigma\ovr2}\<UM\dag+MU\dag\>
\label{PFeff}
\eeq
where the integration is over all smooth fields taking values in $SU(N_{\!f})$,
and standard chiral perturbation theory (XPT) to order $O(p^2)$ has been used%
\footnote{For a discussion in the framework of generalized chiral perturbation
theory see \cite{DescotesStern}.}.
In the representation (\ref{PFeff}) $U\!=\!\exp(\mathrm{i}\sqrt{2}\,\phi/F)$
is the Goldstone manifold and $M\!=\!\mathrm{diag}(m,\ldots,m)\!>\!0$ the quark
mass matrix; $\<\ldots\>$ means a flavour trace, and $F$ is the pion decay
constant to $O(p^2)$.
The key observation by Leutwyler and Smilga is that if they take not only
the box length sufficiently large for XPT to apply%
\footnote{$\Lambda_\mathrm{XPT}$ denotes a QCD intrinsic low-energy scale,
e.g.\ $\Lambda_\mathrm{QCD},\sqrt{\si},F_\pi$.}
but also the quark masses sufficiently light
\beq
{1\ovr\Lambda_\mathrm{XPT}}\ll L \ll {1\ovr M_\pi}
\label{LScondition}
\eeq
the collective (constant mode) excitations are no longer suppressed, but
provide the dominant contribution \cite{GLthermodynamics}, i.e.\ (\ref{PFeff})
is approximately given by
\beq
Z\sim\int dU_0\;e^{V\Sigma\<U_0M\dag+MU_0\dag\>/2}
\label{LSglobalmode}
\eeq
where the path integral has collapsed into a simple group integral given by
the Haar measure.
From (\ref{LSglobalmode}) one gets the contribution of the topological sector
$\nu$ to the partition function in the standard way by substituting
$M\to M\,e^{\mathrm{i}\th/N_{\!f}}$ and Fourier transforming from $\th$ to
$\nu$, whence
\bea
Z_\nu
&\sim&
\int_0^{2\pi} d\th\int dU_0\;e^{-\mathrm{i}\nu\th}\;
e^{V\Sigma\<U_0\,e^{-\mathrm{i}\th/N_{\!f}}M\dag+\mathrm{h.c.}\>/2}
\nonumber
\\
&\sim&
\int_{U(N_{\!f})} d\til U\;(\det\til U)^\nu\;
e^{V\Sigma\<\til UM\dag+\mathrm{h.c.}\>/2}
\label{LSfreeze}
\eea
where in the last line $U_0$ and the factor $e^{-\mathrm{i}\th/N_{\!f}}$ have
been combined into the $U(N_{\!f})$ matrix $\til U$.
From (\ref{LSfreeze}) it follows that, if $V\Sigma m$ is large, the dominant
contribution to the group integral will come from the area where $\til U$ is
in the vicinity of the identity matrix, which means that the determinant is
approximatively one and hence $Z_\nu$ is independent of $\nu$.
A refined analysis shows that this is true for $\nu^2\!\ll\!V\Sigma m/N_{\!f}$,
and the overall distribution is \cite{LeutwylerSmilga}
\beq
Z_\nu\sim e^{-{\nu^2\ovr2\<\nu^2\>}}
\quad
\mathrm{with}
\quad
\<\nu^2\>={V\Sigma m\ovr N_{\!f}}
\qquad\qquad(x\gg1)
\;.
\label{LSgaussian}
\eeq
I shall attach a few comments:

({\sl i\/})
It is worth noticing that the LS-classification of small, intermediate and
large box volumes ($x\!\ll\!1, x\!\simeq\!1, x\!\gg\!1$) does not coincide
with the usual classification on the lattice
($L\!\ll\!M_\pi^{-1}, L\!\simeq\!M_\pi^{-1},L\!\gg\!M_\pi^{-1}$).
Note also that one involves the masses of the {\em sea\/}-quarks, the other
of the {\em current\/}-quarks.

({\sl ii\/})
The analysis by Leutwyler and Smilga covers the symmetry restoration regime
($x\!\ll\!1$) and, on the other side, the regime where SSB is manifest, but the
Goldstone bosons overlap the box ($x\!\gg\!1, L\!\ll\!M_\pi^{-1}$).
The latter condition ---~besides not being useful because it prevents%
\footnote{Unless the current-quark mass is taken considerably {\em heavier\/}
than the sea-quark mass.}
extraction of pionic observables~--- is, in fact, completely inaccessible in
present days simulations: Putting numbers into eqn.~(\ref{LScondition}) one
gets
\beq
\Lambda_\mathrm{XPT}\!\simeq\!200\,\mathrm{MeV}
\quad\Longrightarrow\quad
L\!\simeq\!3\,\mathrm{fm}
\quad\Longrightarrow\quad
M_\pi\!\simeq\!20\,\mathrm{MeV}
\label{LSnumbers}
\eeq
which amounts to a {\em sea\/}-quark mass%
\footnote{Since $M_\pi^2\sim m$, a sea-pion which is lighter than the physical
pion by a factor 7 amounts to a sea-quark mass 49 times smaller than the
phenomenological value of $(m_u\!+\!m_d)/2$. For quark masses the usual
conventions ($\overline{\mathrm{MS}}, \mu\!=\!2\,\mathrm{GeV}$) are adopted.}
of the order of $70\,\mathrm{keV}$.
From this we conclude that, in order to be useful in dynamical QCD simulations,
the result (\ref{LSgaussian}) must turn out to rely exclusively on the
condition $x\!\gg\!1$ being fulfilled, not on the r.h.s.\
of~(\ref{LScondition}).
In other words, our hope is that the latter turns out to be a purely technical
condition, immaterial to the result~(\ref{LSgaussian}).

({\sl iii\/})
Comparing the announcement of the results by Leutwyler and Smilga in the
introduction to their considerations as sketched above, one might worry because
the latter aims at the relative weight of the topological sectors in the
{\em partition function\/}, whereas the introduction mentioned possible
$\nu$-dependencies of {\em observables\/}.
From a formal point of view one may argue that this is not an issue, because
the LS-analysis goes through if the theory is coupled to tiny external sources,
and infinitesimal sources are sufficient to produce Green's functions and
observables.
Hence one expects that a strong dependence or approximate independence of the
partition function w.r.t.\ $\nu$ transfers into a sensitivity or immunity of
correlation functions and observables,
but still one would desire to see this explicitly in numerical data.


\section{Adaptation to 2-flavour QED(2)}

The aim is to provide simulation data in the three LS-regimes $x\!\ll\!1$,
$x\!\simeq\!1$ and $x\!\gg\!1$, while in addition fulfilling the usual
condition $L\!\gg\!1/M_\pi$, i.e.\ the {\em opposite\/} of the r.h.s.\ of
(\ref{LScondition}).

A toy theory where the LS-issue can be studied at greatly reduced costs (in
terms of CPU power) while maintaining all essential features is the massive
multi-flavour Schwinger model, i.e.\ QED(2) with $N_{\!f}\!\geq\!2$ light
degenerate fermions.
The physics of this theory resembles in many aspects QCD(4) slightly {\em above
the chiral phase transition\/}, i.e.\ the chiral condensate vanishes upon
taking the chiral limit and the expectation value of the Polyakov loop is real
and positive.
Furthermore, the required formal analogy holds true:
Like in QCD the gauge fields in the continuum version of the Schwinger model
quantized on the torus fall into topologically distinct classes
\cite{JoosSachsWipf}, i.e.\ they can be attributed a topological index.
The latter agrees with the number of left- minus right-handed zero-modes of the
massless Dirac operator, i.e.\ the index theorem holds true
\cite{JoosSachsWipf}.

There is, however, an (apparent) problem: As we have seen in the previous
section, the physics in the large LS-regime ($x\!\gg\!1$) is governed by the
onset of SSB for the global $SU(N_{\!f})_A$ flavour-group (though the box
volume is still finite), but it is well known that in 2 dimensions no SSB with
the associate production of Goldstone bosons takes place \cite{Coleman}.

The reason why this apparent deficiency proves immaterial relates to the fact
that the multi-flavour Schwinger model (with $N_{\!f}$ {\em massless\/}
flavours) shows a second order phase transition with critical temperature
$T_c\!=\!0$, a result due to Smilga and Verbaarschot \cite{SmilgaVerbaarschot}.
The theory is most appropriately seen as the limiting case of an extension
where the axial flavour symmetry is explicitly broken, e.g.\ by tiny quark
masses.
In the strong-coupling limit%
\footnote{Note that in 2 dimensions the charge has the dimension of a mass.}
$m\!\ll\!g$ the spectrum is found to involve a heavy particle with mass
\cite{SegreWeisberger}
\beq
M_+\simeq\sqrt{N_{\!f}}{g\ovr\sqrt{\pi}}+O(m)
\label{SMetaprime}
\eeq
and $N_{\!f}^2\!-\!1$ light particles with mass \cite{Coleman, Smilga92}
\beq
M_-\simeq\mathrm{const}\;g^{1/(N_{\!f}\!+\!1)}\,m^{N_{\!f}/(N_{\!f}\!+\!1)}
\;.
\label{SMquasigold}
\eeq
The state (\ref{SMetaprime}) is often called a ``massive photon'', but it is
more appropriately seen as the analogue of the $\et'$ in QCD, as it is the
lightest flavour singlet and stays massive in the chiral limit.
The states (\ref{SMquasigold}) are ``{\em Quasi\/}-Goldstones'' and we shall
call them ``pions'', but it is important to keep in mind that they differ from
the usual ``{\em Pseudo\/}-Goldstones'' (as encountered in QCD) as they do not
turn into true Goldstone bosons upon taking $m\!\to\!0$; they rather become
{\em sterile\/} if the chiral limit%
\footnote{Analogous phenomena are observed, if the axial flavour symmetry is
broken by boundary conditions rather than quark masses: The chiral condensate
(i.e.\ the ``would be order parameter'') vanishes with a power-law dependence
as the spatial box length is sent to infinity at zero temperature, but
exponentially fast if the temperature is non-zero \cite{SMboundaryconditions}.}
is performed.
That conforms with Coleman's theorem \cite{Coleman} which forbids the
existence of massless interacting particles in 2 dimensions.
The point I shall stress is this: As long as the axial flavour symmetry is
explicitly broken ---~which, in the following, is true, as sea-quarks are taken
massive~--- the ``Quasi-Goldstones'' are, for practical purposes, as good as
``Pseudo-Goldstones''; the difference shows only up upon taking the chiral
limit.
Hence even the large LS-regime, where long-range Green's functions are
dominated by pionic excitations, finds its analogue in the (massive)
multi-flavour Schwinger model.

One more practical issue has to be resolved: The original definition
(\ref{LSPdef}) of the LS-parameter $x$ involves the quantity $\Sigma$, i.e.\
(the absolute value of) the chiral condensate in the chiral limit.
The latter, however, is exactly zero, due to Coleman's theorem.
A practical way out of this is based on Smilga's observation
that upon combining (\ref{SMquasigold}) with the expression \cite{Smilga92}
\beq
|\<\psb\ps\>|\simeq\mathrm{const}\;
g^{2/(N_{\!f}\!+\!1)}\,m^{(N_{\!f}\!-\!1)/(N_{\!f}\!+\!1)}
\label{SMcond}
\eeq
for the chiral condensate one gets the relation \cite{Smilga92}
\beq
m\,|\<\psb\ps\>|\simeq\,{0.388\ovr(2.008)^2}\,M_\pi^2
\label{GORinMSM}
\eeq
where I have already plugged in the nonuniversal (i.e.\ $N_{\!f}$-dependent)
constants for the 2-flavour case \cite{Smilga97}.
Relation (\ref{GORinMSM}) being the QED(2) analogue of the
Gell-Mann--Oakes--Renner PCAC-relation
\beq
m\Sigma\simeq{1\ovr2}M_\pi^2F_\pi^2
\qquad\qquad(\forall N_{\!f})
\label{GORinQCD}
\eeq
in QCD means that I may {\em define\/} the LS-parameter in the lattice studies
presented below as
\beq
x\equiv{0.388\ovr(2.008)^2}\,VM_\pi^2
\qquad\qquad(N_{\!f}=2)
\label{LSPinMSM}
\eeq
as one may rewrite it in the case of QCD (to leading order in XPT)
\beq
x={1\ovr2}\,VM_\pi^2F_\pi^2
\qquad\qquad(\forall N_{\!f})
\label{LSPinQCD}
\eeq
in a way which involves just physical degrees of freedom%
\footnote{Note that in order to determine $x$ in a lattice study through
(\ref{LSPinQCD}) one has to plug in $M_\pi$ and $F_\pi$ of the
{\em sea\/}-pion, i.e.\ in practice of a pion constructed from current-quarks
having exactly the same mass as their sea-partners.}.

In order to investigate the LS-issue, I have chosen to implement the Schwinger
model with a pair of (dynamical) staggered fermions, using the Wilson gauge
action $S_{\rm gauge}\!=\!\beta\sum(1\!-\cos\theta_\Box\!)$.
Since the staggered Dirac operator is represented by relatively small matrices
(see below), I have decided to compute the determinants exactly, using the
routines ZGEFA and ZGEDI from the LINPACK package.

I compare the three regimes $x\!\ll\!1, x\!\simeq\!1, x\!\gg\!1$ to each other
using three dedicated simulations:
I have chosen to vary the box volume at fixed $\beta\!=\!1/(ag)^2\!=\!3.4$ and
fixed staggered quark mass $m=0.09$ (everything in lattice units).
The three regimes are represented by the three volumes
$V\!=\!8\!\times\!4, V\!=\!18\!\times\!6, V\!=\!40\!\times\!10$, where always
periodic boundary conditions in the first (spatial) direction and thermal
boundary conditions in the second (euclidean timelike) direction have been
used.
The LS-parameter (\ref{LSPinMSM}) takes the values $x\!\simeq\!0.33,
x\!\simeq\!1.12, x\!\simeq\!4.16$ respectively, while the pion (pseudo-scalar
iso-triplet) has a (common) mass%
\footnote{The formula used is the prediction (\ref{SMquasigold}), which seems
adequate since $m/g\!\simeq\!0.166\!\ll\!1$.}
$M_\pi\!=\!2.008\cdot0.542^{1/3}\,0.09^{2/3}\!=\!0.329$ and hence a correlation
length $\xi_\pi\!=\!3.04$ as to fit into the box
(almost) in all three regimes -- as is usual in a lattice context.
Each time 3200 decorrelated (w.r.t.\ the plaquette) configurations have been
generated.

For the type of investigation I am aiming at configurations must be assigned
an index $\nu$. I have implemented both the geometric definition
$\nu_{\rm geo}\!=\!{1\over2\pi}\sum\log U_\Box$
and the field theoretic definition $\nu_{\rm fth}\!=\!\kappa\,\nu_{\rm nai},
\nu_{\rm nai}\!=\!\sum\sin\theta_\Box$ with the renormalization factor
$\kappa\!\simeq\!1/(1-\langle S_{\rm gauge}\rangle/\beta V)$ \cite{LueSmiVin}.
A configuration is assigned an index only if the geometric and the
field-theoretic definition, after rounding to the nearest integer, agree.
This turned out to be the case for 3197, 3163, 2818 configurations, i.e.\ on
a 99.9\%, 98.8\%, 88.1\% basis on the small/intermediate/large lattice.
This fraction being so high means that in practice an assignment can be done
{\em without cooling\/} for the overwhelming majority of configurations.
The remaining ones are just not assigned an index; they are used to compute
the unseparated observable (e.g.\ the ``complete'' heavy quark potential; see
below), the latter, however, turns out to be unaffected by whether they are
included or not.

\begin{figure}
\epsfig{file=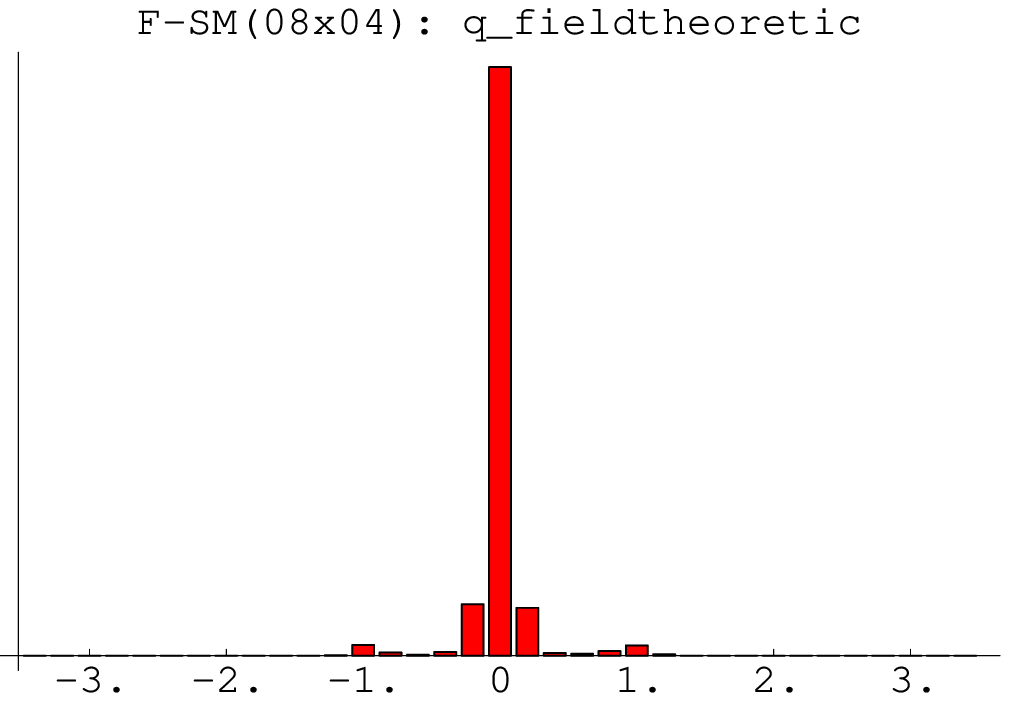,
height=4.4cm,width=6.4cm,angle=0}
\hfill
\epsfig{file=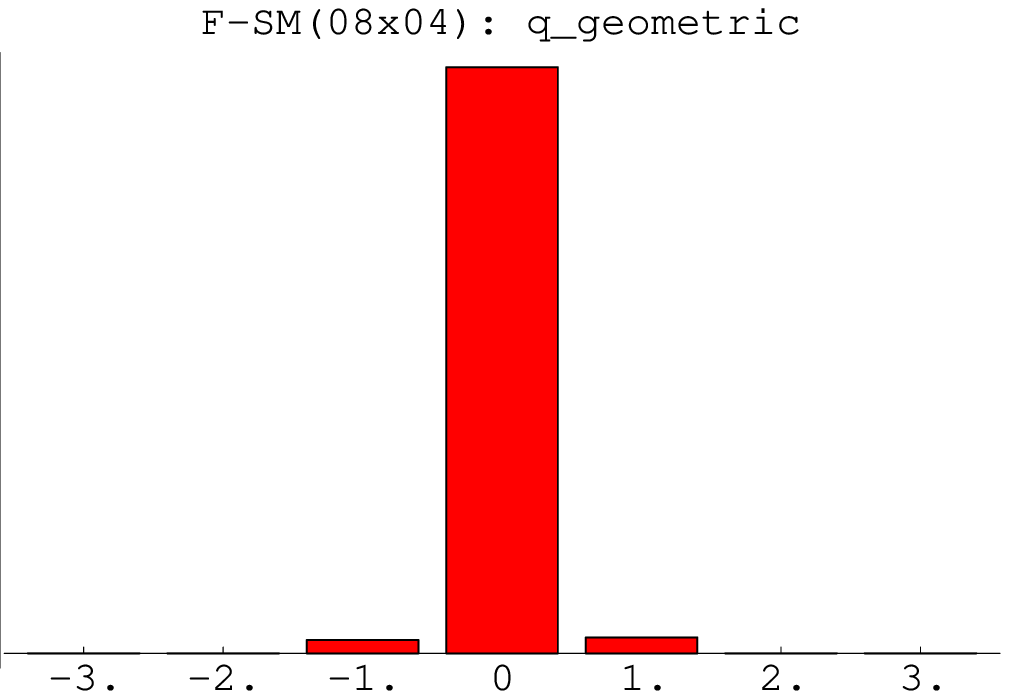,
height=4.4cm,width=6.4cm,angle=0}
\vspace*{2mm}
\\
\epsfig{file=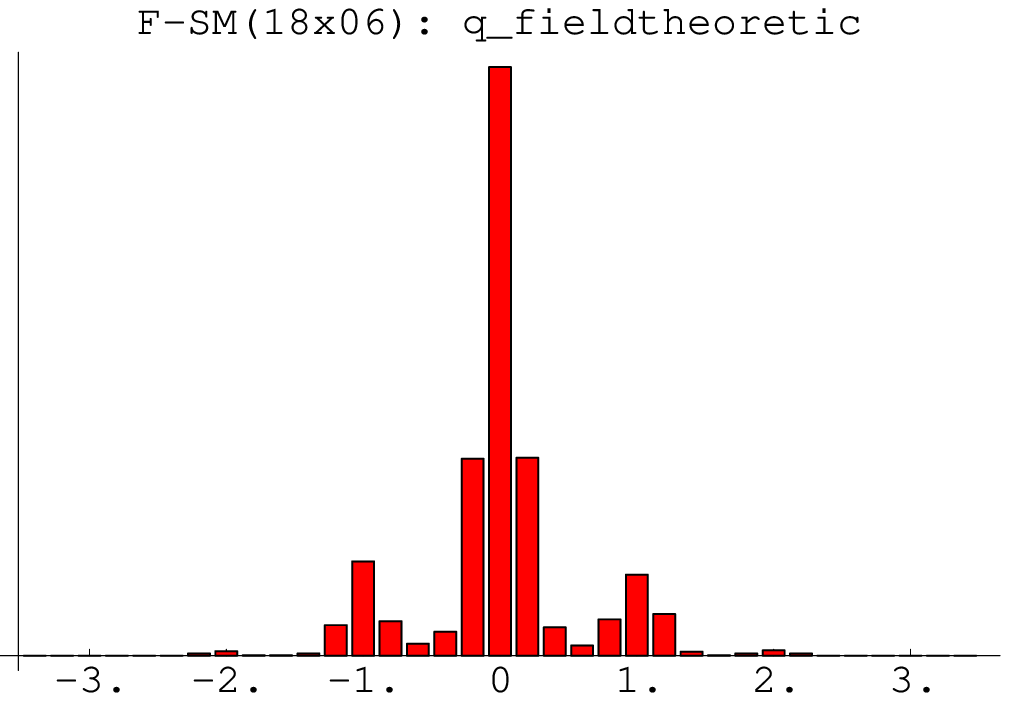,
height=4.4cm,width=6.4cm,angle=0}
\hfill
\epsfig{file=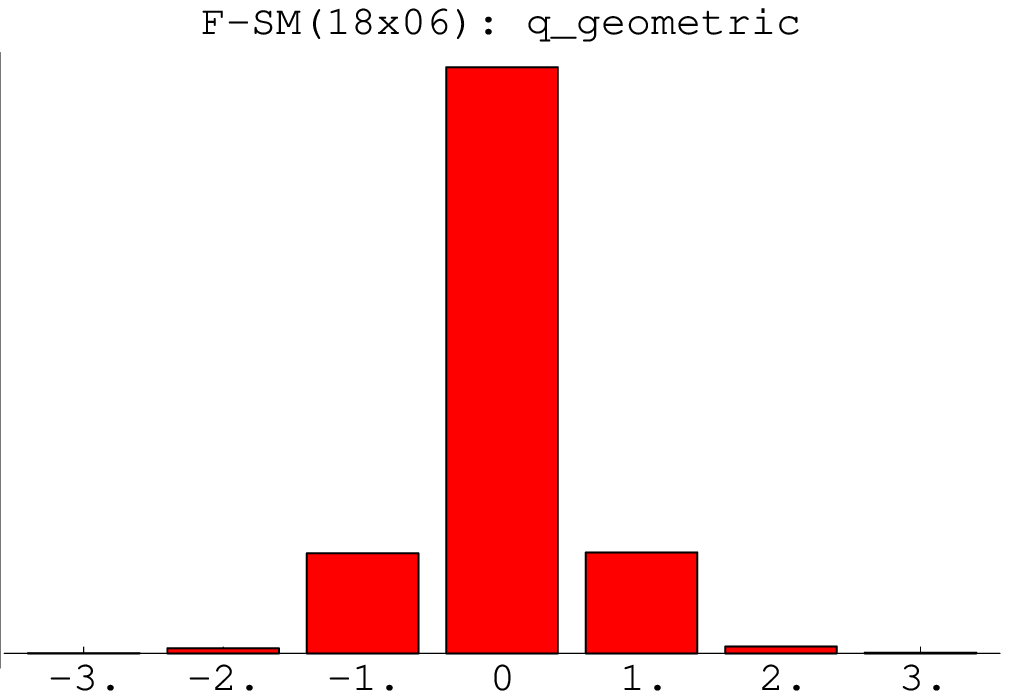,
height=4.4cm,width=6.4cm,angle=0}
\vspace*{2mm}
\\
\epsfig{file=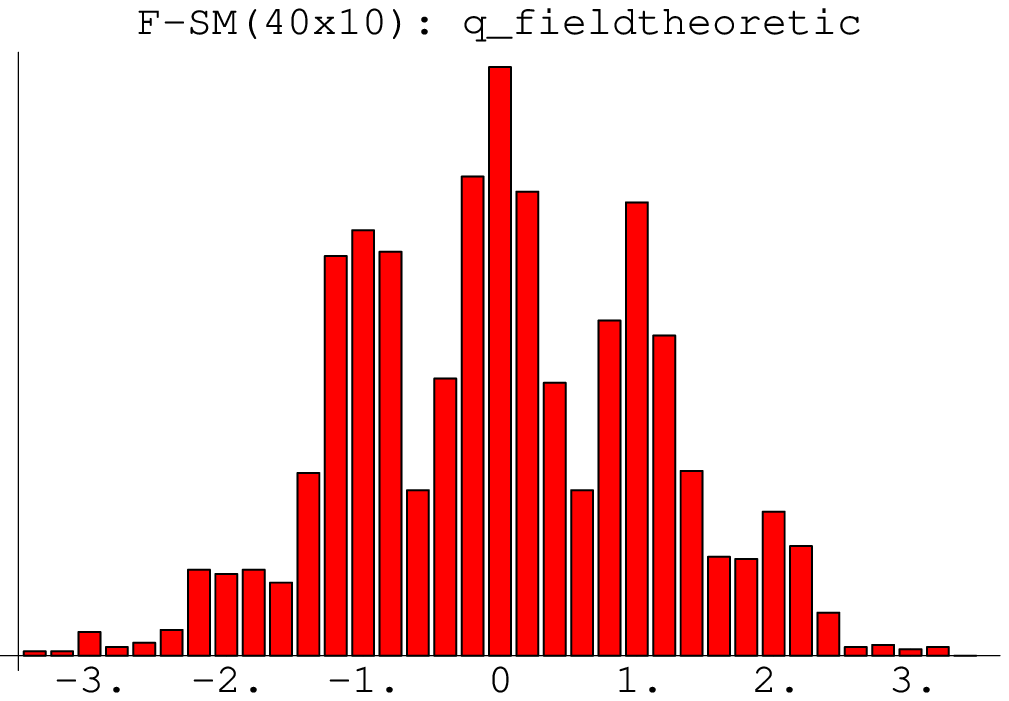,
height=4.4cm,width=6.4cm,angle=0}
\hfill
\epsfig{file=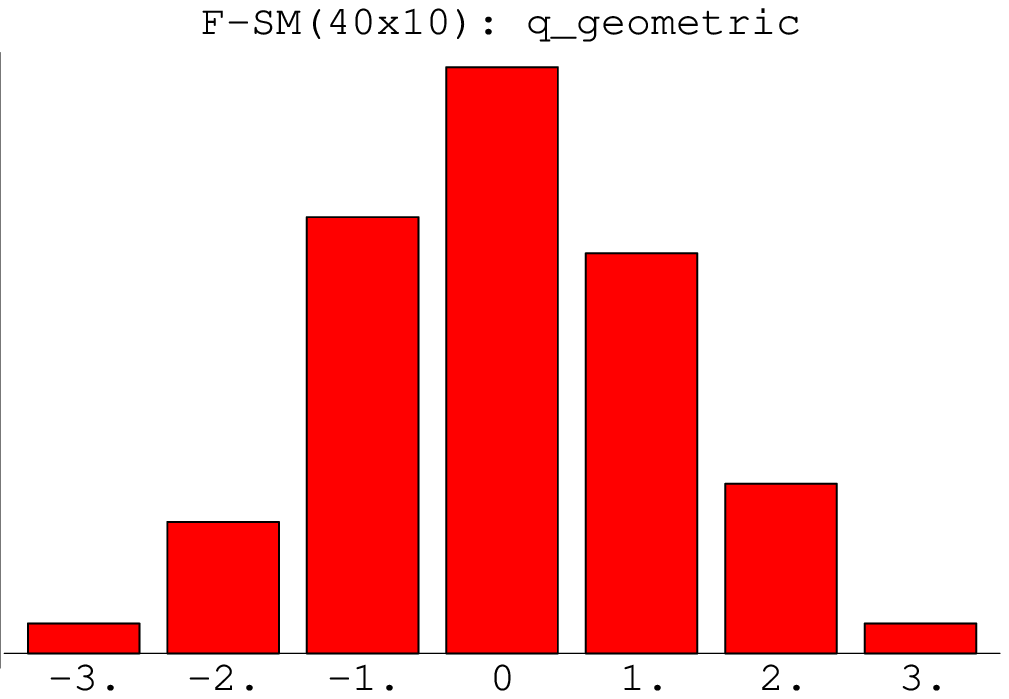,
height=4.4cm,width=6.4cm,angle=0}
\vspace*{-2mm}
\caption{\sl\small
Distribution of\/ $\nu_\mathrm{fth}$ (left) and $\nu_\mathrm{geo}$ (right) in
the small, intermediate and ``large'' Leutwyler-Smilga regimes, respectively.}
\end{figure}

Having simulation data which are supposed to reflect the situation in the
small/intermediate/large LS-regimes respectively, one may want to check the
overall distribution of topological charges as encountered in the three runs.
As one can see from the r.h.s.\ of Fig.~2, the overall distribution is
noticeably different in the three regimes.
The predictions by Leutwyler and Smilga for the two extreme cases $(x\!\ll\!1$
and $x\!\gg\!1$) seem to be well reproduced: For small $x$ the partition
function $Z$ (and hence the sample) is dominated by $Z_0$, the contribution
from the topologically trivial sector, whereas for large $x$ the distribution
gets broad and seems compatible with a gaussian.
It is worth mentioning that the latter holds true even though the pion did not
overlap the box; rather the usual relation $\xi_\pi\!\ll\!L$ applied.
This is a first indication that the r.h.s.\ of (\ref{LScondition}) in the
LS-analysis might indeed be a purely technical condition, immaterial to what
they claim, namely that in the large $x$ regime the topological charge is
``an irrelevant concept'' or, in other words, that physical observables do not
depend on $\nu$ if $x\!\gg\!1$ \cite{LeutwylerSmilga}.


\section{Sectoral reweightings and renormalizations}

Since the LS-issue is peculiar to the full (unquenched) theory, an attempt to
understand by which mechanism the three regimes differ from each other may lead
one to investigate how the functional determinant or specifically its
contribution to the total action per continuum-flavour
\beq
S_\mathrm{fermion}=-\log(\det(D\!\!\!\!/+\!m))+\mathrm{const}
\label{Sfermion}
\eeq
relates to the contribution from the gauge field and, in addition, how such a
relationship might depend on the topological charge of the background.
The idea is thus to study such a relationship {\em sectorally\/}, i.e.\ after
the complete sample has been separated into subsamples with a fixed topological
charge.
Since the theory (at $\th\!=\!0$) is P/CP-symmetric, the configurations from
the sectors $\pm\nu$ may be combined into a single subsample characterized by
$|\nu|$.

\begin{figure}
\begin{tabular}{lcr}
\epsfig{file=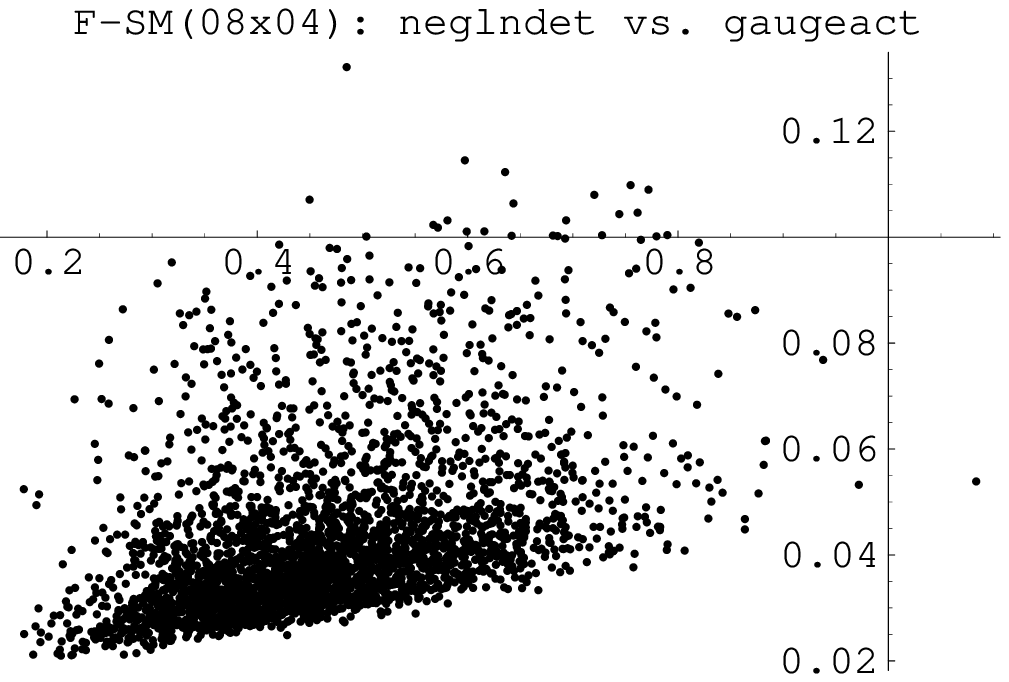,
height=3.4cm,width=4.2cm,angle=90}&
\epsfig{file=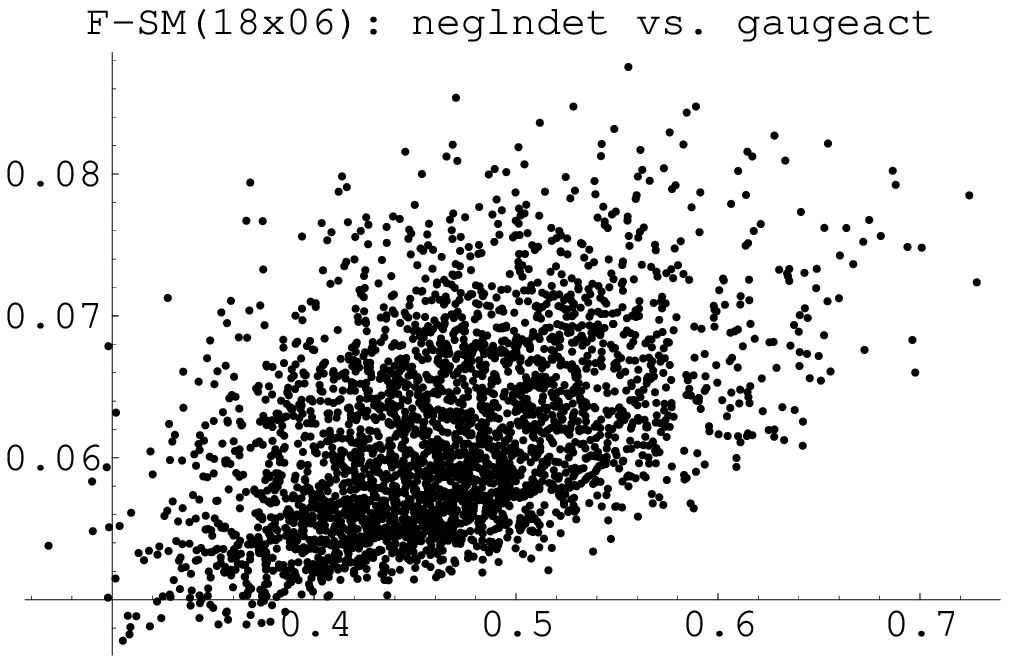,
height=3.4cm,width=4.2cm,angle=90}&
\epsfig{file=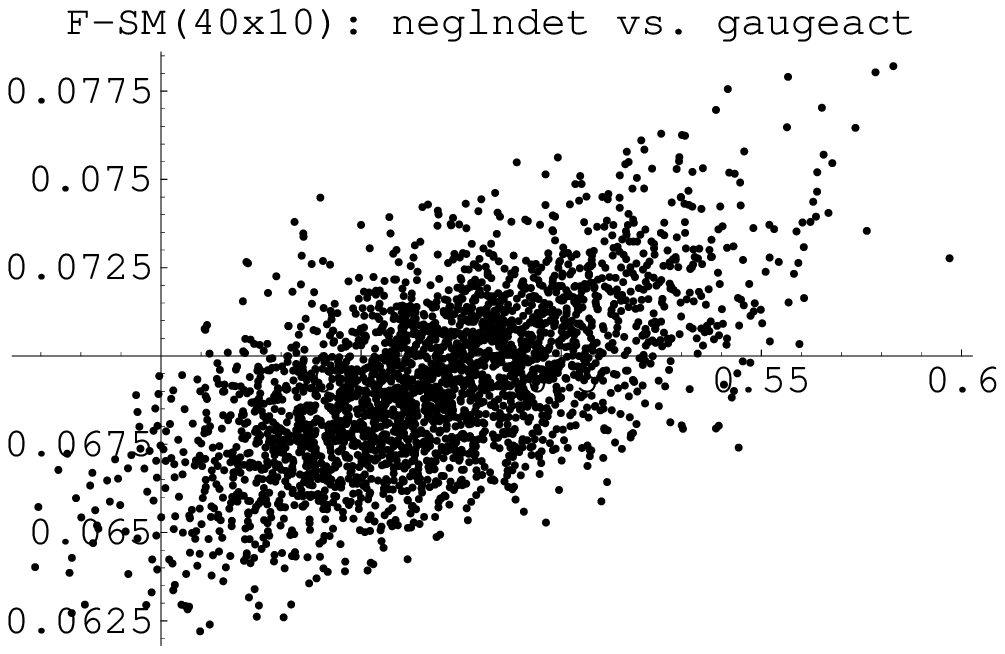,
height=3.4cm,width=4.2cm,angle=90}
\\
\hspace*{40mm}&
\hspace*{40mm}&
\hspace*{6mm}
\epsfig{file=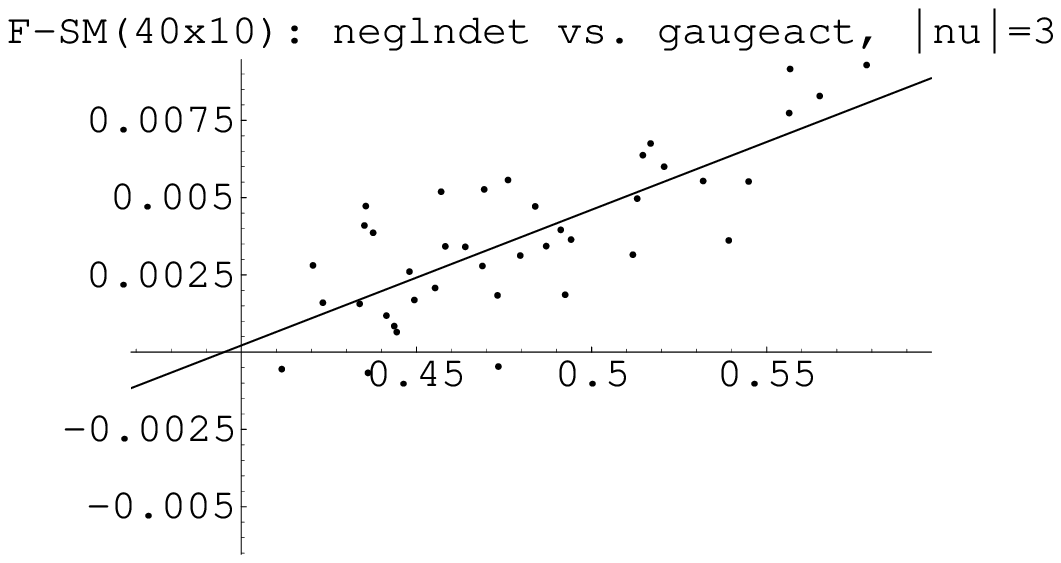,
height=3.4cm,width=4.2cm,angle=90}
\\
{}&
\epsfig{file=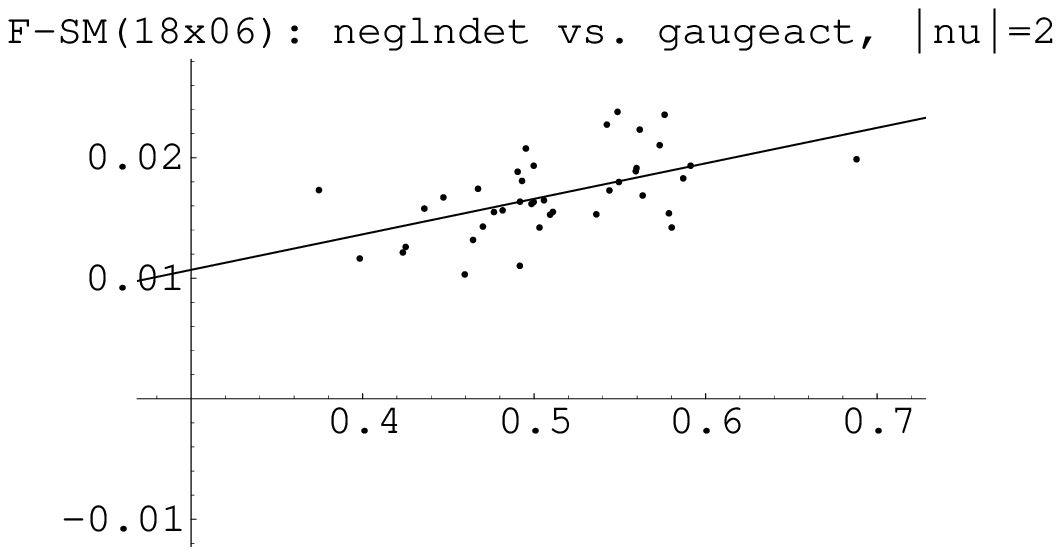,
height=3.4cm,width=4.2cm,angle=90}&
\epsfig{file=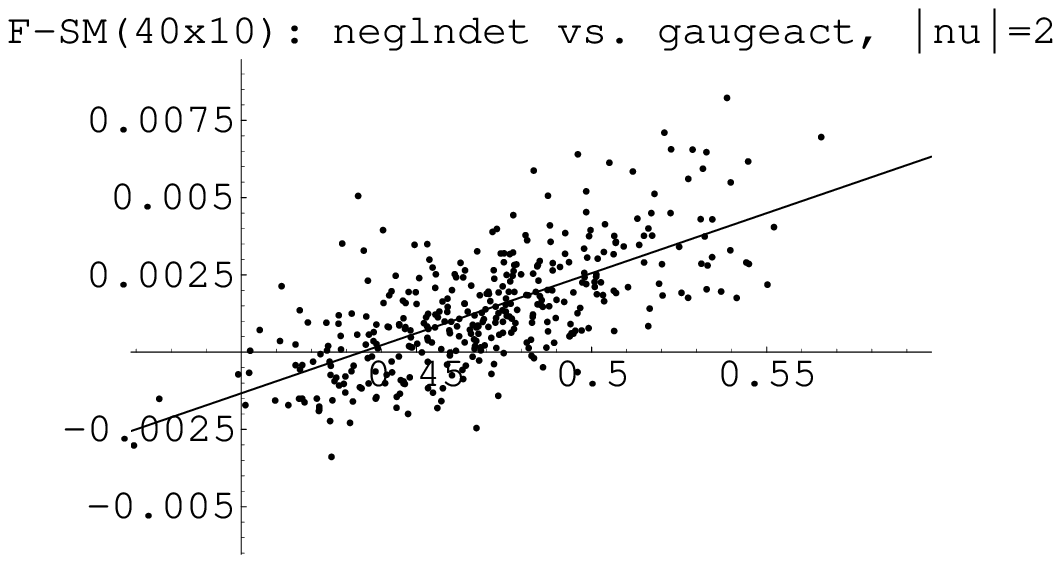,
height=3.4cm,width=4.2cm,angle=90}
\\
\epsfig{file=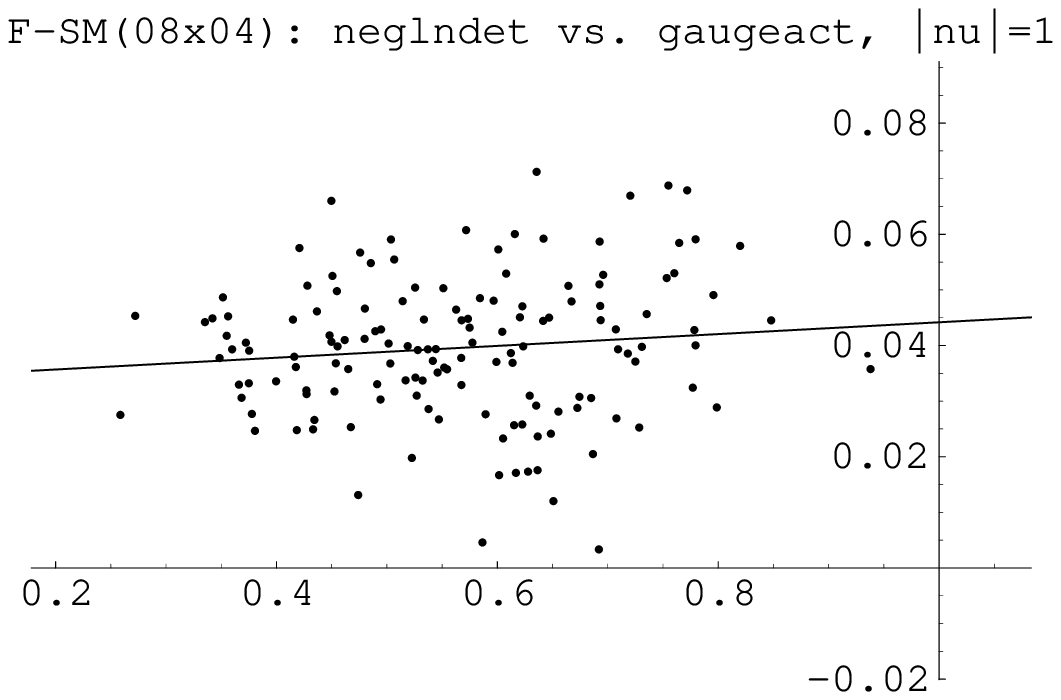,
height=3.4cm,width=4.2cm,angle=90}&
\epsfig{file=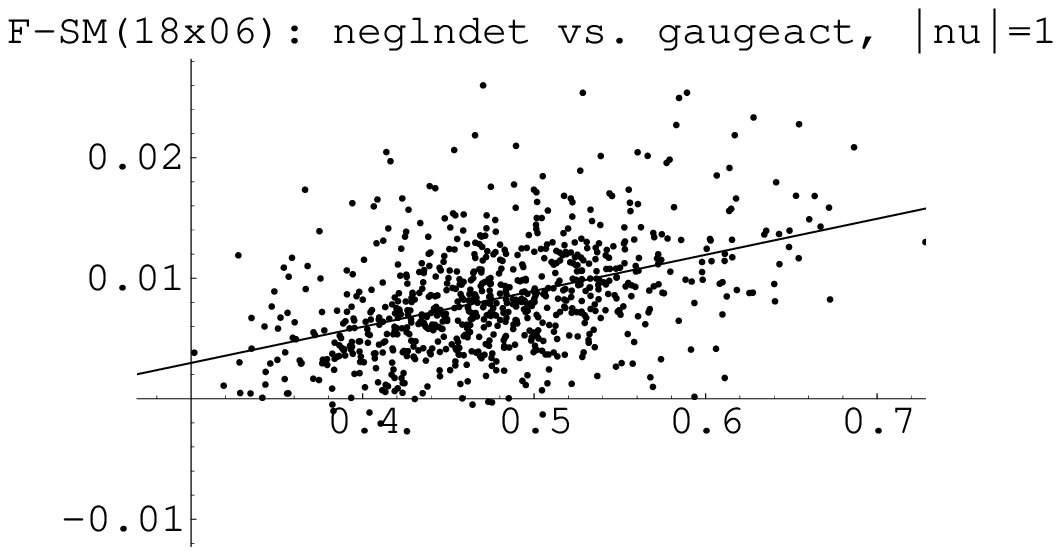,
height=3.4cm,width=4.2cm,angle=90}&
\epsfig{file=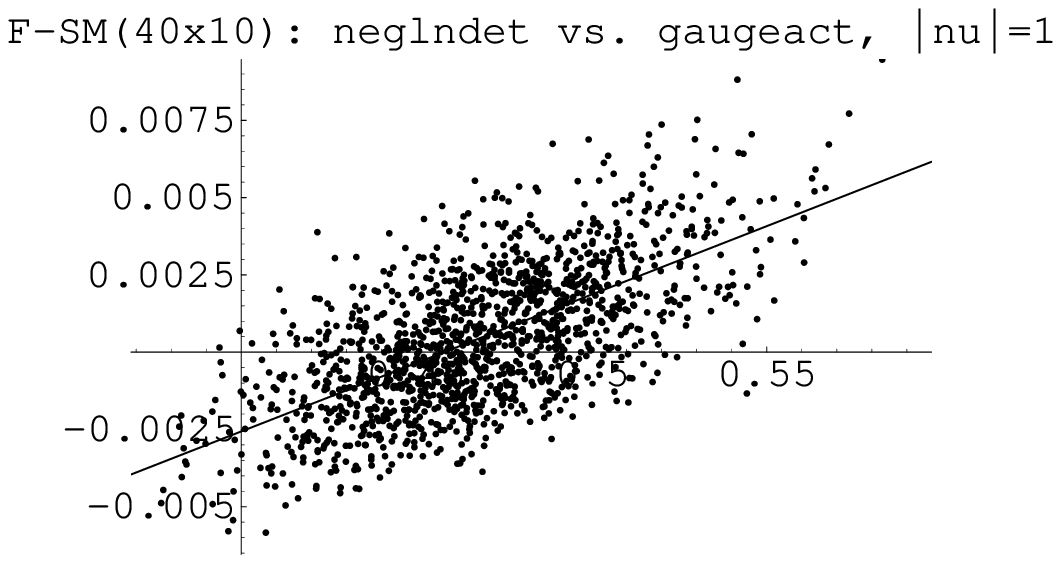,
height=3.4cm,width=4.2cm,angle=90}
\\
\epsfig{file=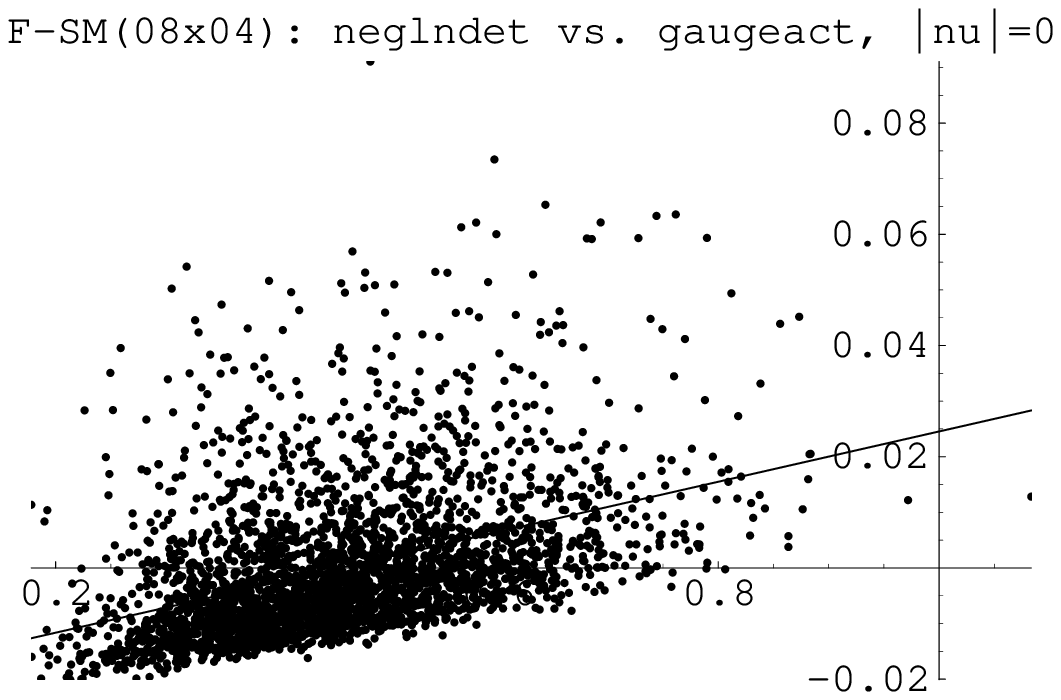,
height=3.4cm,width=4.2cm,angle=90}&
\epsfig{file=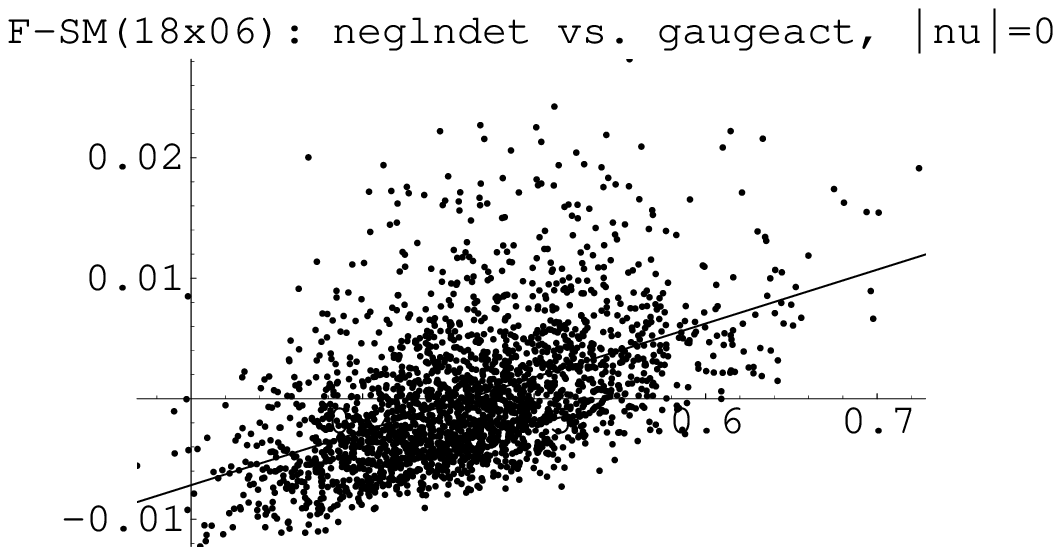,
height=3.4cm,width=4.2cm,angle=90}&
\epsfig{file=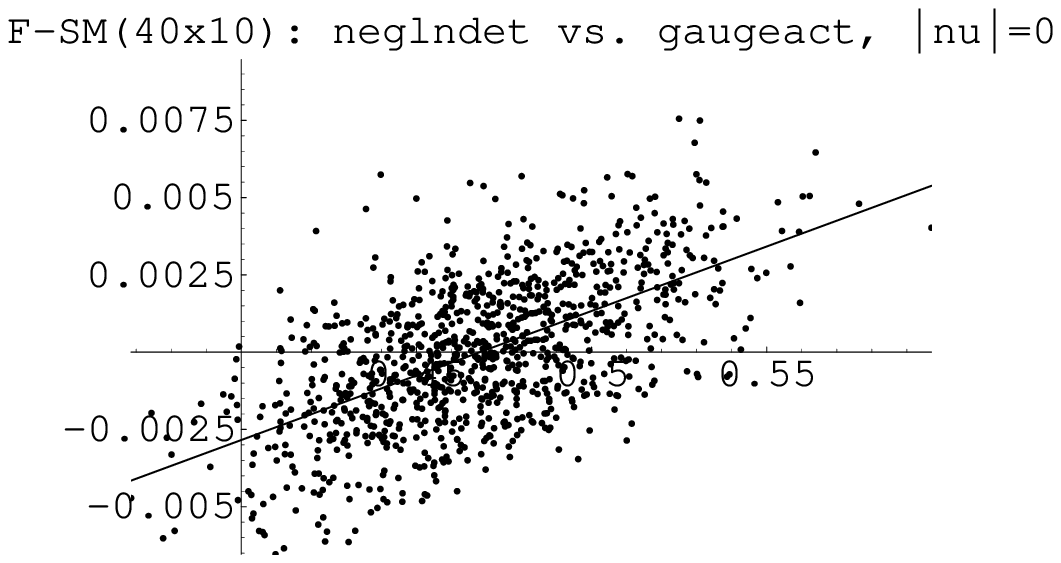,
height=3.4cm,width=4.2cm,angle=90}
\end{tabular}
\vspace*{-3mm}
\caption{\sl\small
Scatter plot of $S_{\rm fermion}/V\!=\!-\!\log(\det(D\!\!\!\!/+\!m))/V$ versus
$S_{\rm gauge}/V$ on the complete sample (rightmost column) and on subsamples
with $|\nu|\!=\!0,1,2,3$ in the small (top), intermediate (middle) and large
(bottom) Leutwyler-Smilga regime. In the sectoral plots the constant in
(\ref{Sfermion}) is chosen such that $\<S_{\rm fermion}\>_{\nu\!=\!0}\!=\!0$.}
\end{figure}

As one can see from the rightmost column of Fig.\ 3, the contribution per
continuum flavour to the total action, $S_{\rm fermion}$, shows a {\em positive
correlation\/} with the gauge action $S_{\rm gauge}$.
This means that, on a qualitative level, switching on the functional
determinant amounts to an {\em increase\/} of $\beta$, which is a well-known
feature in (full) QCD too.
However, the correlation is weak, i.e.\ the ``cigars'' in the unseparated
plots are thick.
A key observation is that the {\em correlation improves, if one separates the
sample into subsamples with fixed\/} $|\nu|$, as is done in Fig.\ 3.
It is worth noticing that in general (i.e.\ without specifying the LS-regime)
the best linear fit for $S_{\rm fermion}$ as a function of $S_{\rm gauge}$
depends (in offset and slope) on $|\nu|$.
This means that, to a first approximation, the functional determinant results
in an {\em overall suppression of higher topological sectors\/} w.r.t\ lower
ones and to a {\em sectorally different renormalization\/} of $\beta$.
As we shall see, these two features constitute the place where the three
LS-regimes differ from each other.

First, we shall consider the offset of the regression line in the
$\nu\!=\!\pm1$ sectors compared to the line in the sector with $\nu\!=\!0$,
as this is a measure how much extra suppression the first topologically
nontrivial sector receives on average compared to the trivial one due to the
functional determinant.
From the first and second column in Fig.\ 3 one sees that this offset (and in
general the offset between neighboring sectors) {\em decreases\/} with
increasing $x$.
This provides an explanation \cite{Gattringeretal} for the broadening of the
$\nu$-distribution inherent to passing from the small to the intermediate or
from the intermediate to the large LS-regime (cf.\ Fig.\ 2).
For $x\!\ll\!1$ the offset $\<S_{\rm fermion}\>_1-\<S_{\rm fermion}\>_0$ is
larger than the fluctuations in $S_{\rm fermion}$ within the zero-charge
sector, and this means means that in the small LS-regime the functional
determinant acts as an almost {\em strict constraint\/} to the topologically
trivial sector.
For $x\!\simeq\!1$ the offset has roughly the same size as typical fluctuations
of $S_{\rm fermion}$ in the sector with $\nu\!=\!0$; as a result of this one
can say that in the intermediate regime the functional determinant acts as a
{\em ``soft constraint''\/} to low-lying $|\nu|$ values.
For $x\!\gg\!1$ finally, $\<S_{\rm fermion}\>_1-\<S_{\rm fermion}\>_0$ is small
compared to the fluctuations in each one of these sectors; hence no such
effective description proves appropriate in the large LS-regime.

Passing on to the slope of the regression line in the correlation
$\<S_{\rm fermion}\>$ versus $\<S_{\rm gauge}\>$, we notice from Fig.\ 3 that
this slope decreases (in general) with increasing $|\nu|$.
This means that the effective renormalization
\beq
\beta^{(\nu)}\;\longrightarrow\;
\beta^{(\nu)}_\mathrm{eff}=(1+N_{\!f}\,\mathrm{slope}(\nu))\;\beta^{(\nu)}
\label{BetaEff}
\eeq
due to the functional determinant is (in general) stronger for sectors with
low absolute value of the net topological charge.
This effect of sectoral dependence of the effective renormalization factor for
$\beta$ diminishes with increasing $x$; eventually, in the large $x$ regime, it
is completely gone.

Thus, the large LS-regime is unique, as the determinant results only in an
overall suppression of higher topological sectors, but not in a sectoral
dependence of $\beta_\mathrm{eff}$ (and hence, in 4 dimensions, of the lattice
spacing in physical units).
This seems consistent with the claim by Leutwyler and Smilga that physical
observables do not depend on $\nu$ in the large $x$ regime, but it is of course
not a sufficient condition for the latter to be true.


\section{Sectoral heavy quark free energies}

A quantity which is easily accessible in lattice studies and which might help
to elucidate the physical meaning of the LS-classification is the Polyakov
loop, i.e.\ the trace of a chain of link variables which winds once around the
torus in the euclidean timelike direction.
Its logarithm represents (up to a factor) the free energy of an external
(``heavy'') quark which is brought into the system without possibility to
influence it through back-reaction.
As in the previous section the analysis shall be done {\em sectorally\/},
i.e.\ on classes of configurations with a fixed value of $|\nu|$.

\begin{figure}
\begin{tabular}{lcr}
\epsfig{file=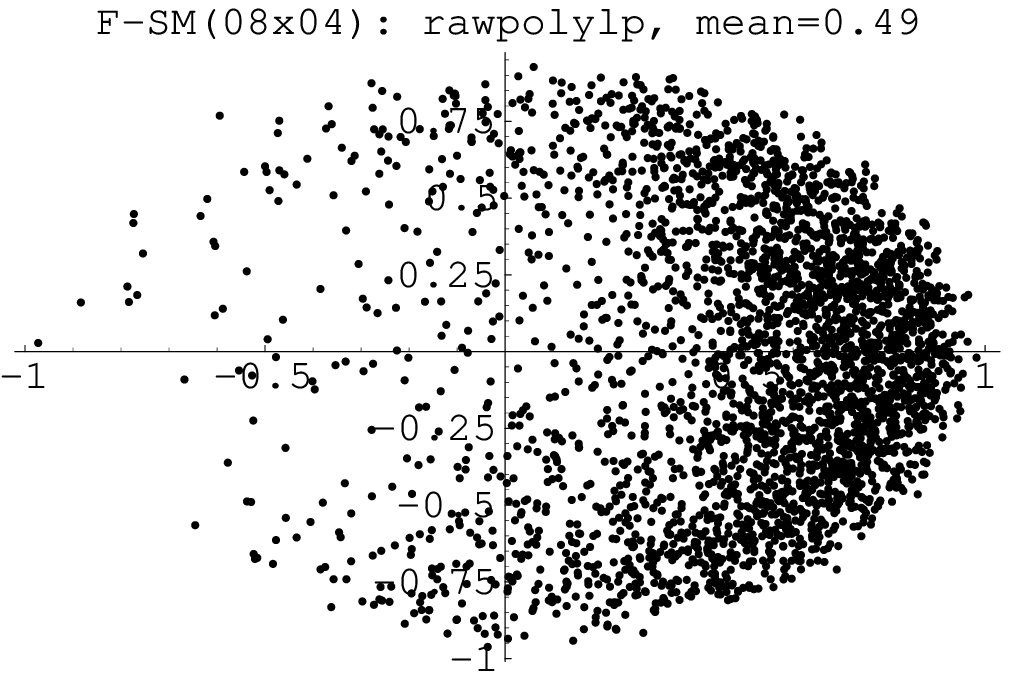,
height=3.4cm,width=4.2cm,angle=90}&
\epsfig{file=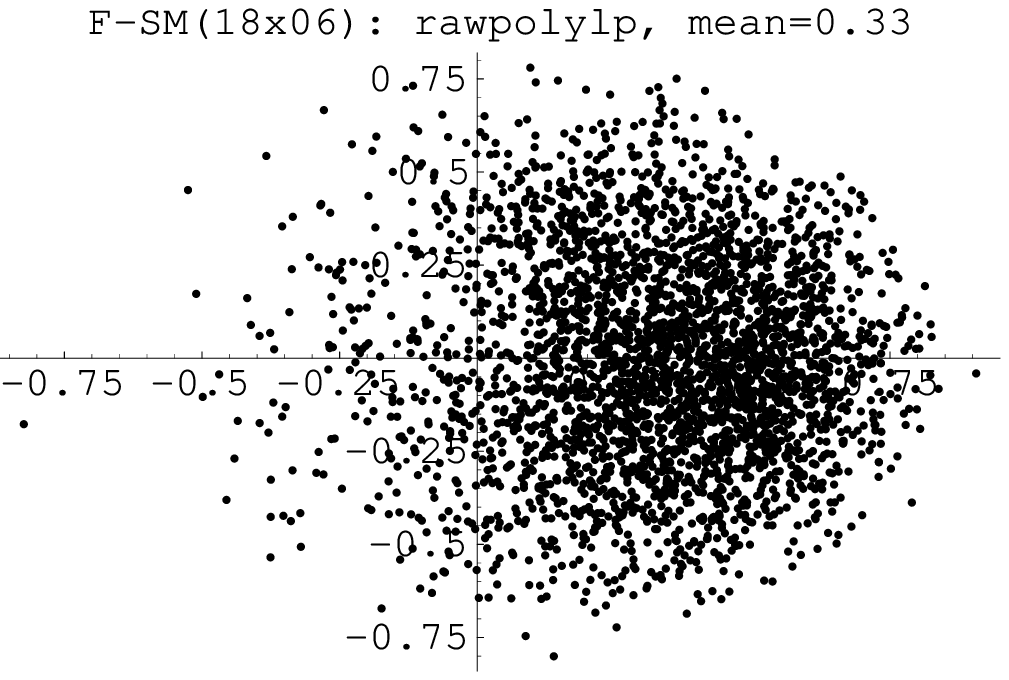,
height=3.4cm,width=4.2cm,angle=90}&
\epsfig{file=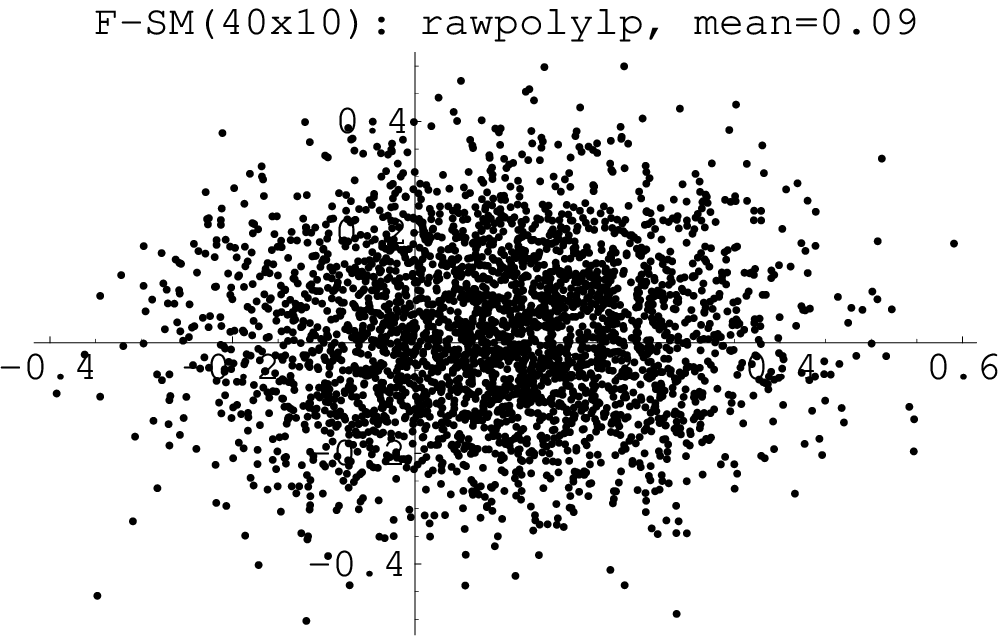,
height=3.4cm,width=4.2cm,angle=90}
\\
\hspace*{40mm}&
\hspace*{40mm}&
\hspace*{6mm}
\epsfig{file=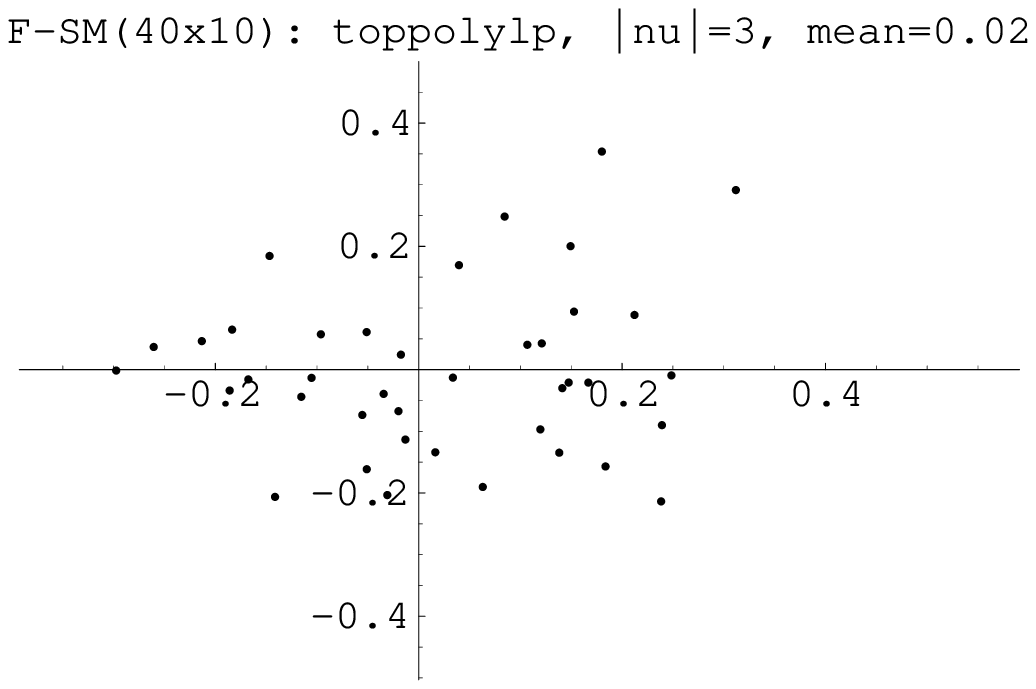,
height=3.4cm,width=4.2cm,angle=90}
\\
{}&
\epsfig{file=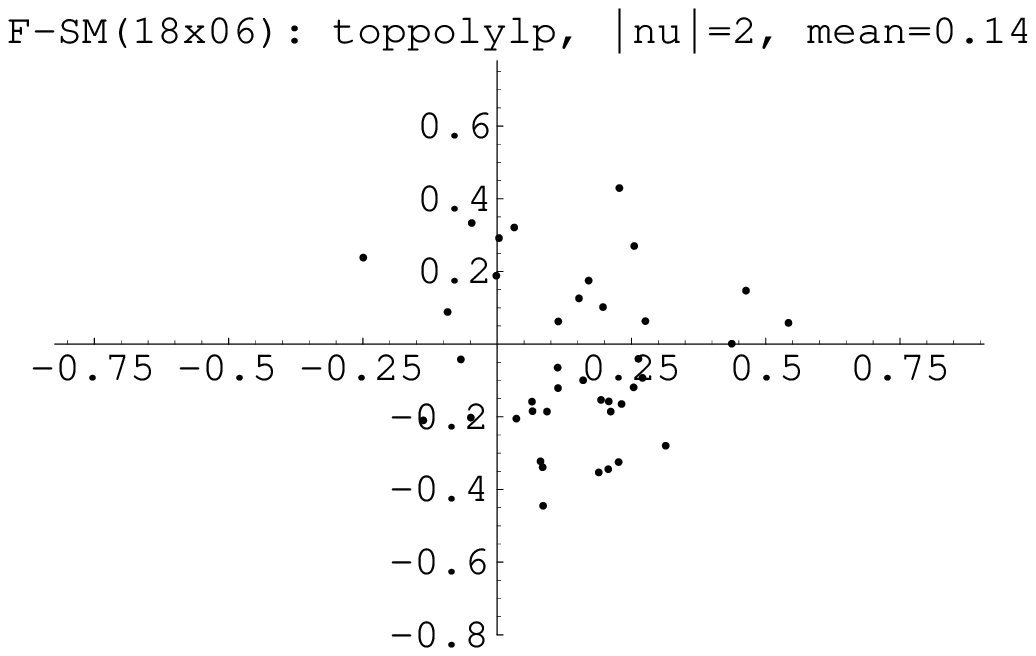,
height=3.4cm,width=4.2cm,angle=90}&
\epsfig{file=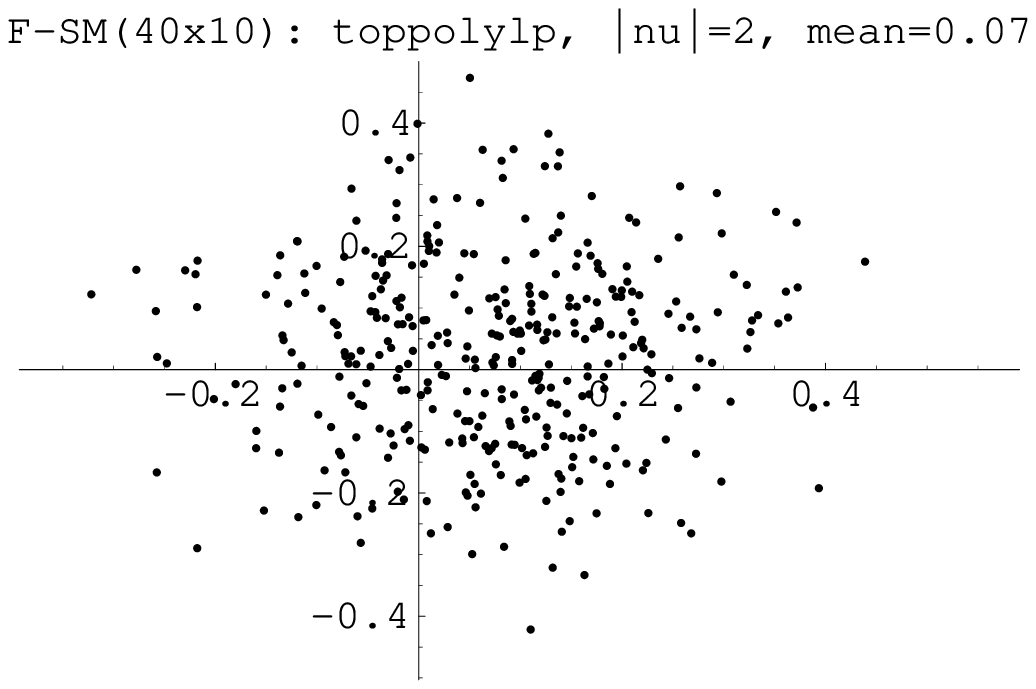,
height=3.4cm,width=4.2cm,angle=90}
\\
\epsfig{file=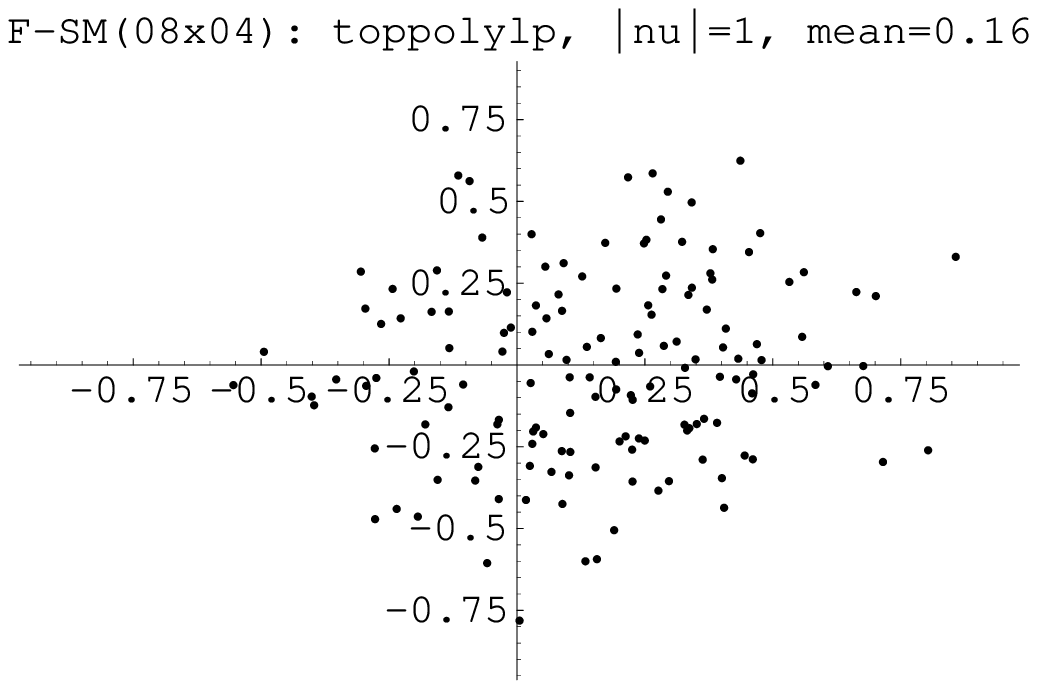,
height=3.4cm,width=4.2cm,angle=90}&
\epsfig{file=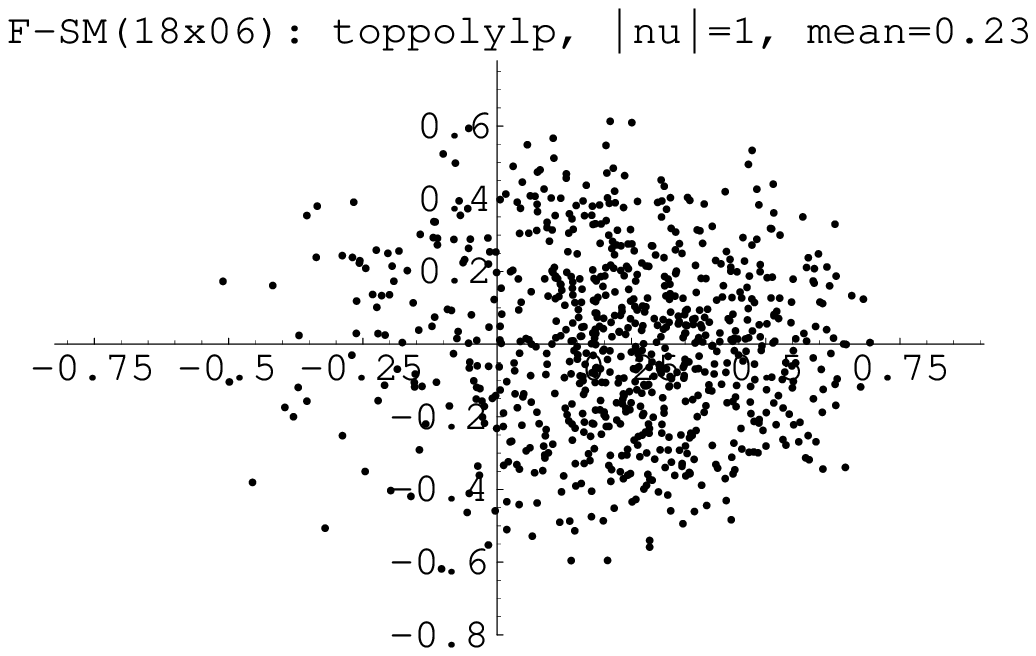,
height=3.4cm,width=4.2cm,angle=90}&
\epsfig{file=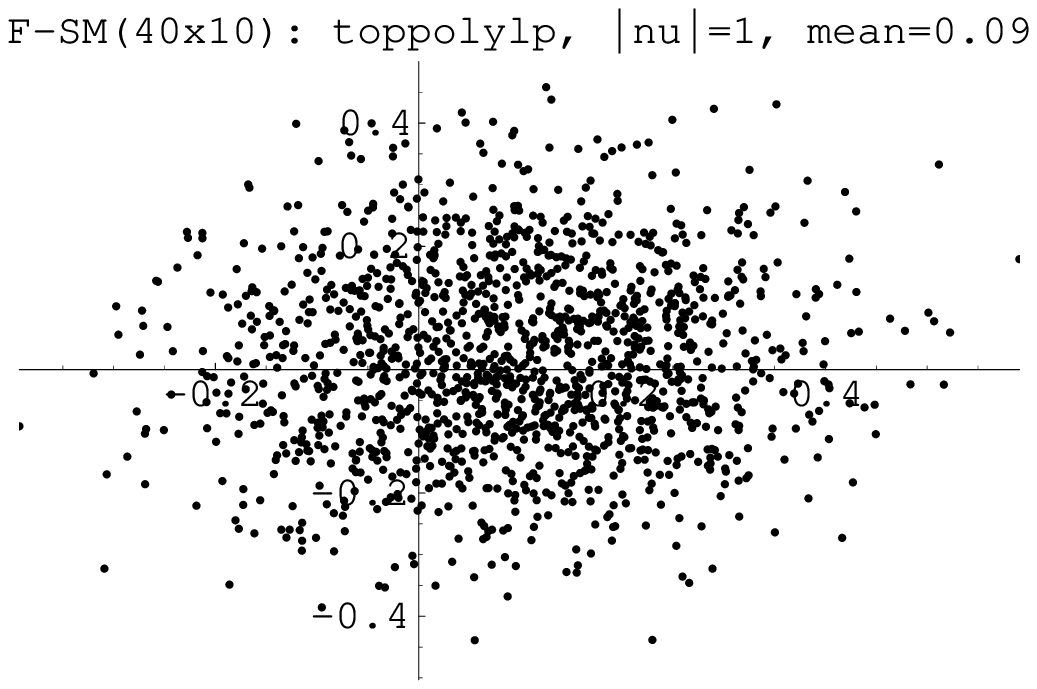,
height=3.4cm,width=4.2cm,angle=90}
\\
\epsfig{file=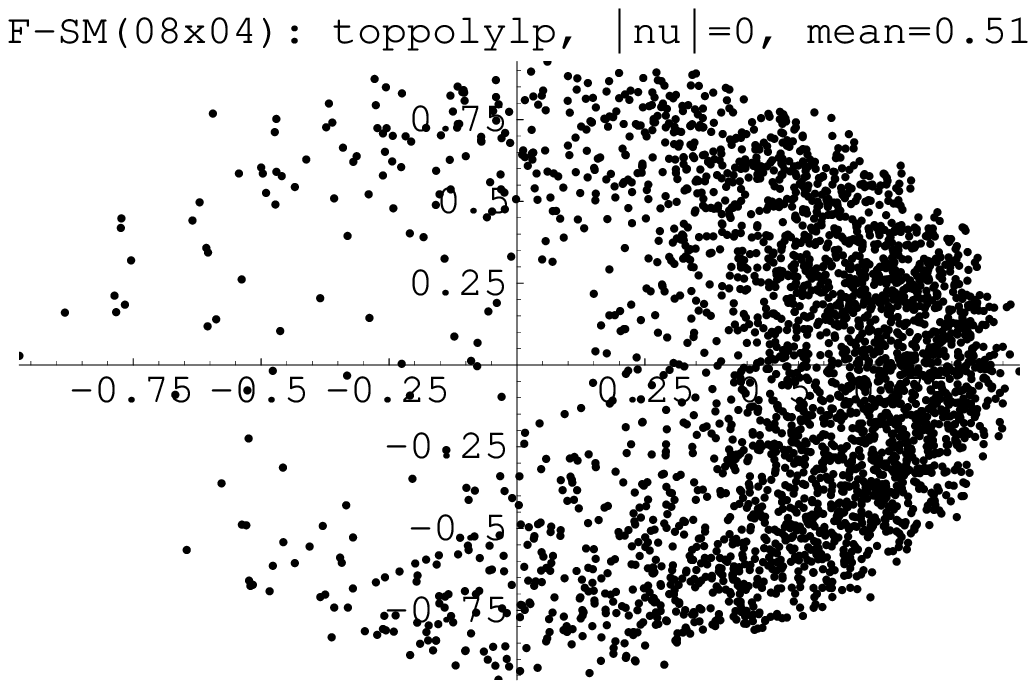,
height=3.4cm,width=4.2cm,angle=90}&
\epsfig{file=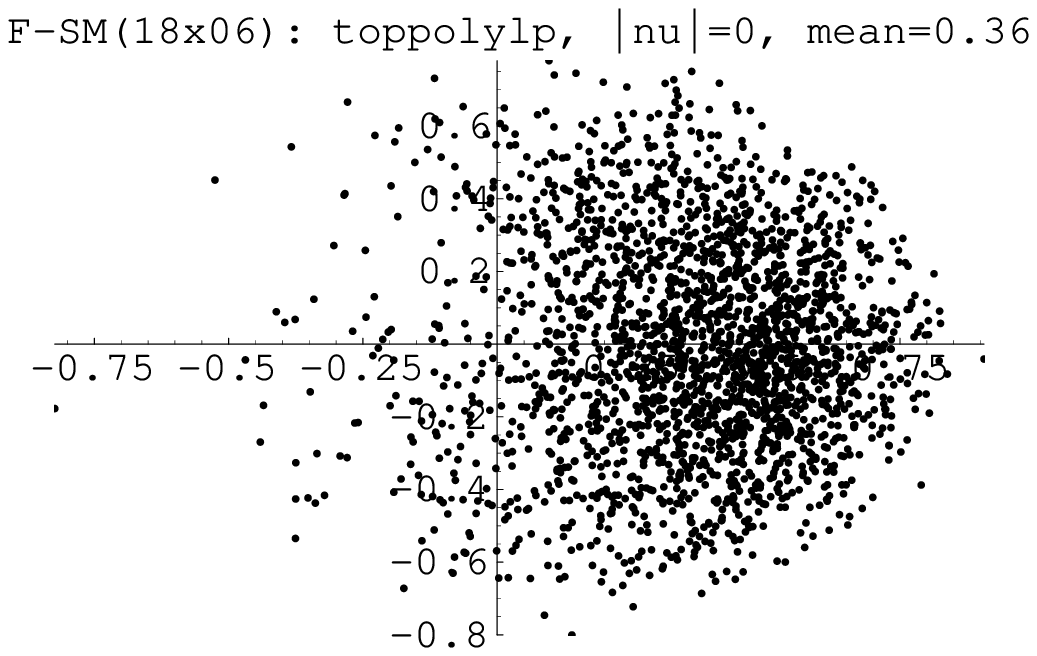,
height=3.4cm,width=4.2cm,angle=90}&
\epsfig{file=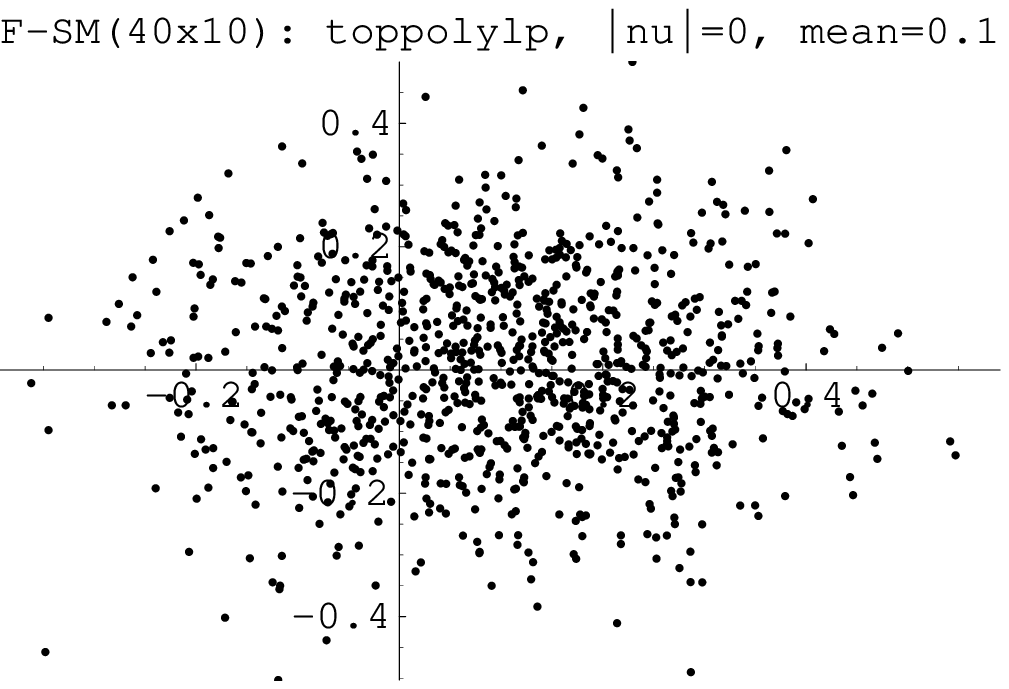,
height=3.4cm,width=4.2cm,angle=90}
\end{tabular}
\vspace*{-2mm}
\caption{\sl\small
Scatter plot of the Polyakov loop on the complete sample (rightmost column) and
on subsamples with $|\nu|\!=\!0,1,2,3$ in the small (top), intermediate
(middle) and large (bottom) Leutwyler-Smilga regime.}
\end{figure}

The result of such an analysis for the Polyakov loop is displayed in Fig.\ 4.
As one can see from the rightmost column, the expectation value $\<L\>$ on
the ``complete'' (unseparated) sample decreases if the box volume is increased.
What is more interesting is the following:
The physical (unseparated) expectation value may be understood as a weighted
mean
\beq
\<L\>=\sum_{|\nu|\ge0}\,p_{|\nu|}\,\<L\>_{|\nu|}
\label{PolyloopWeighted}
\eeq
of sectoral expectation values $\<L\>_{|\nu|}$, where the factor $p_{|\nu|}$
reflects the combined weight of the sectors $\pm\nu$ in the histogram on the
r.h.s.\ of Fig.~2.
From the first line in Fig.\ 3 one notices that the sectoral expectation
values for $|\nu|\!=\!0,1$ in the small LS-regime differ quite drastically on
the scale set by the physical expectation value, i.e.\
$|\<L\>_{|0|,\,x\ll1}\!-\!\<L\>_{|1|,\,x\ll1}|/\<L\>_{x\ll1}\simeq0.7$.
In the intermediate regime the analogous ratios for {\em neighboring sectors\/}
(i.e.\ $|\<L\>_{|0|,\,x\simeq1}\!-\!\<L\>_{|1|,\,x\simeq1}|/\<L\>_{x\simeq1}$
and $|\<L\>_{|1|,\,x\simeq1}\!-\!\<L\>_{|2|,\,x\simeq1}|/\<L\>_{x\simeq1}$)
happen to be smaller, but comparing the highest to the lowest available
$|\nu|$, one finds again
$|\<L\>_{|0|,\,x\simeq1}\!-\!\<L\>_{|2|,\,x\simeq1}|/\<L\>_{x\simeq1}\simeq0.7$.
Finally, in the large LS-regime, there is an almost perfect consistency between
sectoral Polyakov loop expectation values for low-lying $|\nu|$ (i.e.\
$|\nu|\in\{0,1,2\}$), but there is no numerical evidence in favour of an
{\em overall consistency\/}; if one builds the analogous ratio for the lowest
and highest $|\nu|$ available, the quantity
$|\<L\>_{|0|,\,x\gg1}\!-\!\<L\>_{|3|,\,x\gg1}|/\<L\>_{x\gg1}$ is found to be
again of order one.

The Polyakov loop is the first example of a physical observable for which we
see a difference between expectation values in {\em neighboring sectors\/} (for
sufficiently low $|\nu|$ values) disappear upon taking the large $x$ limit.
However, even in the large LS-regime no supporting evidence has been found that
$\<L\>_\nu$ might be {\em completely independent\/} of $\nu$.
It seems that $\<L\>_{|\nu|}$ proves sensitive on topology {\em even in the
large LS-regime\/} when jumping from $|\nu|_\mathrm{min}$ to
$|\nu|_\mathrm{max}$.


\section{Sectoral heavy quark potentials}

The second observable which has been evaluated for each sector separately is
the heavy quark potential as determined from the correlator of two Polyakov
loops --- see Fig.\ 5 and Tabs.\ 1-3.
The first thing to notice is that in each regime the physical potential
(the one in the rightmost column in Fig.~5) bends to the right.
This curvature is a direct indication of the {\em screening brought by the
dynamical fermions\/}; the quenched potential is strictly linear%
\footnote{Unlike in QCD(4) there is no $1/r$ short-distance tail; the
one-photon exchange contributes a linear piece in 2 dimensions.}.

\begin{figure}
\begin{tabular}{lcr}
\epsfig{file=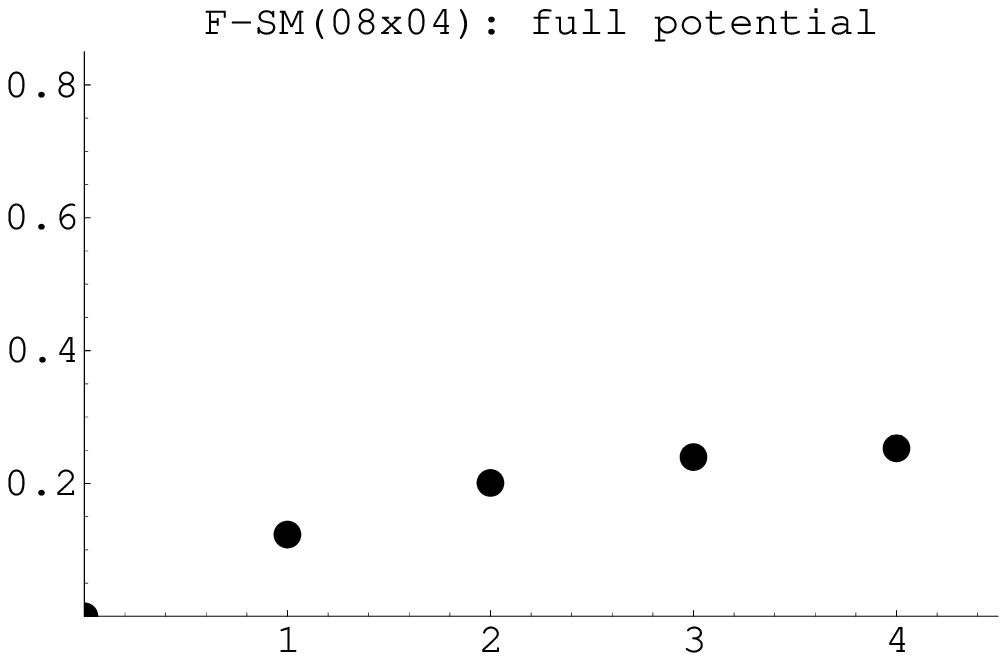,
height=3.4cm,width=4.2cm,angle=90}&
\epsfig{file=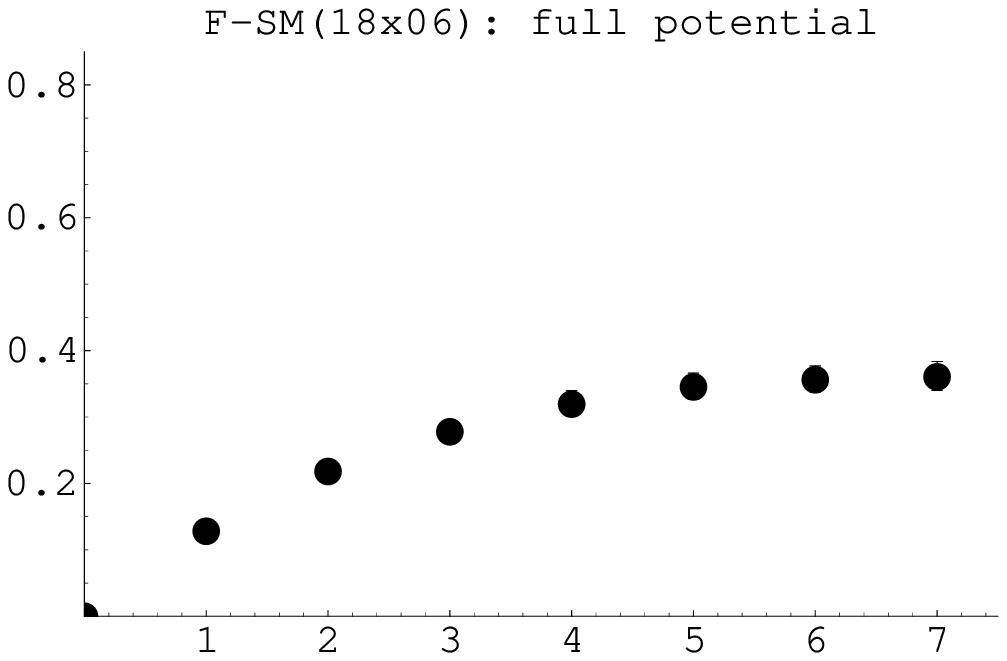,
height=3.4cm,width=4.2cm,angle=90}&
\epsfig{file=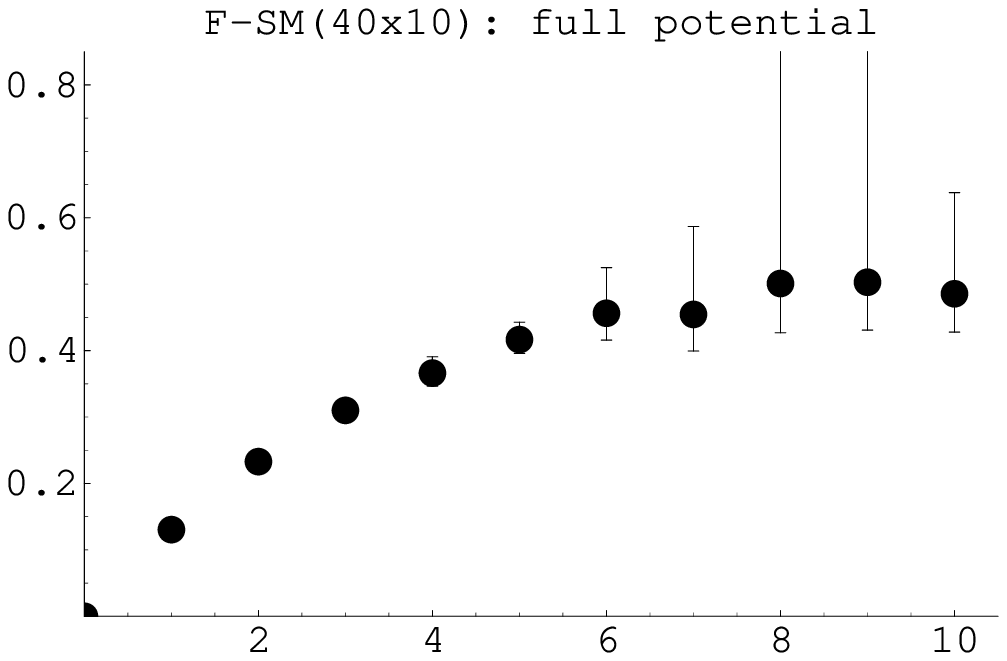,
height=3.4cm,width=4.2cm,angle=90}
\\
\hspace*{40mm}&
\hspace*{40mm}&
\hspace*{6mm}
\epsfig{file=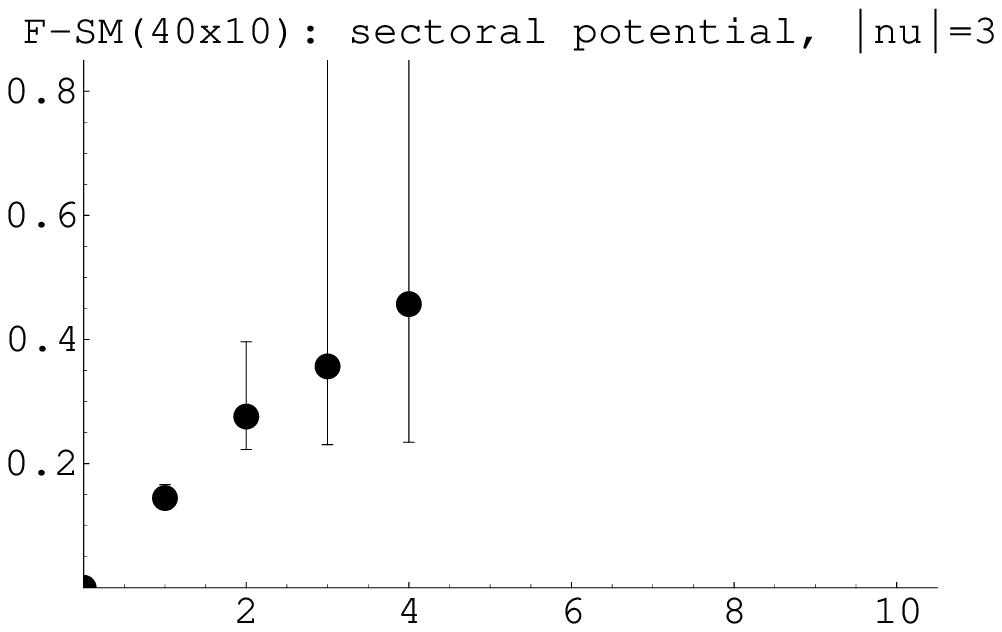,
height=3.4cm,width=4.2cm,angle=90}
\\
{}&
\epsfig{file=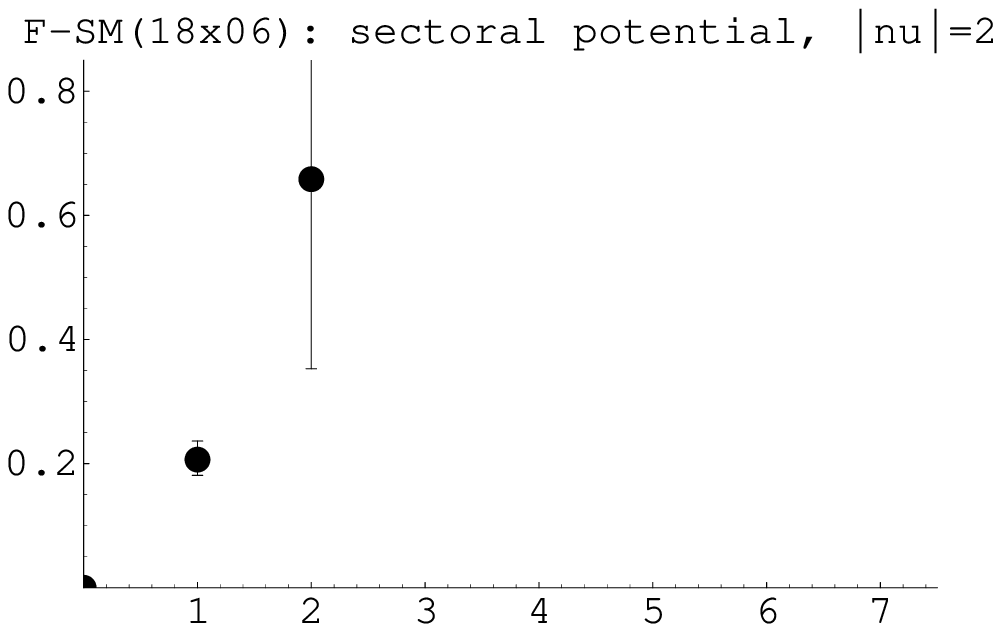,
height=3.4cm,width=4.2cm,angle=90}&
\epsfig{file=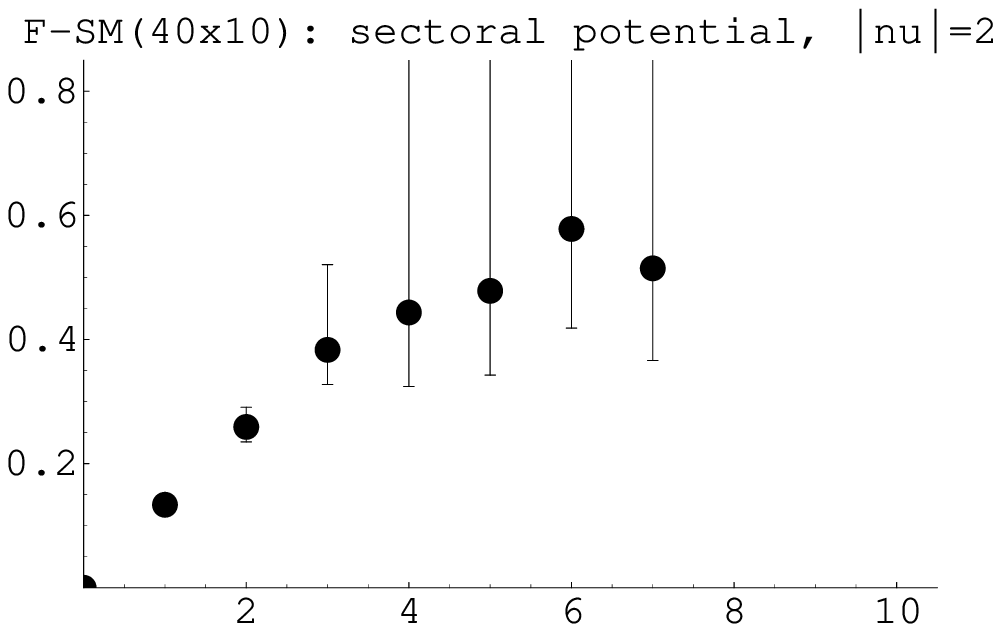,
height=3.4cm,width=4.2cm,angle=90}
\\
\epsfig{file=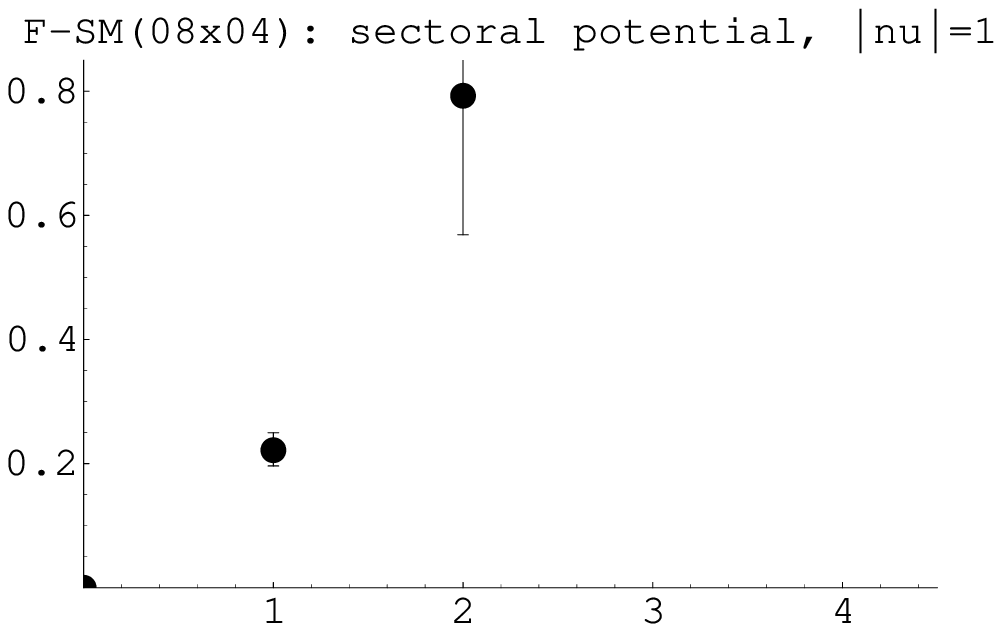,
height=3.4cm,width=4.2cm,angle=90}&
\epsfig{file=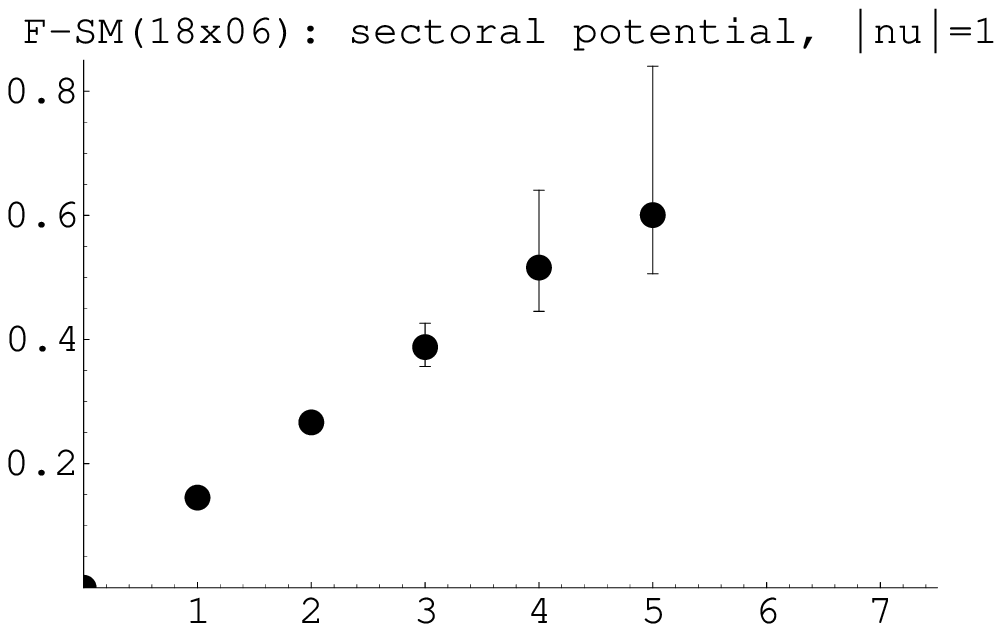,
height=3.4cm,width=4.2cm,angle=90}&
\epsfig{file=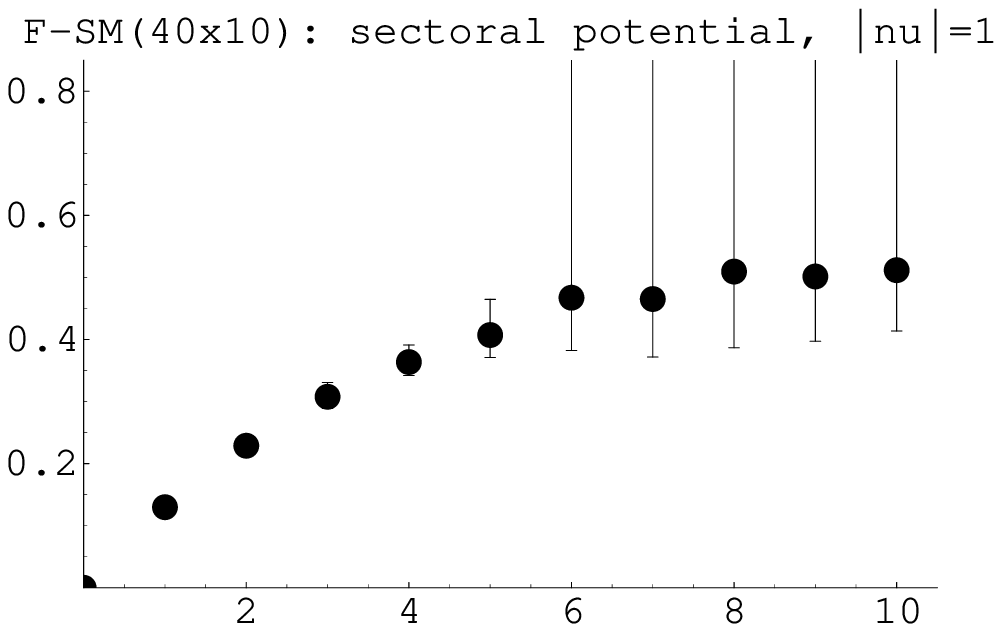,
height=3.4cm,width=4.2cm,angle=90}
\\
\epsfig{file=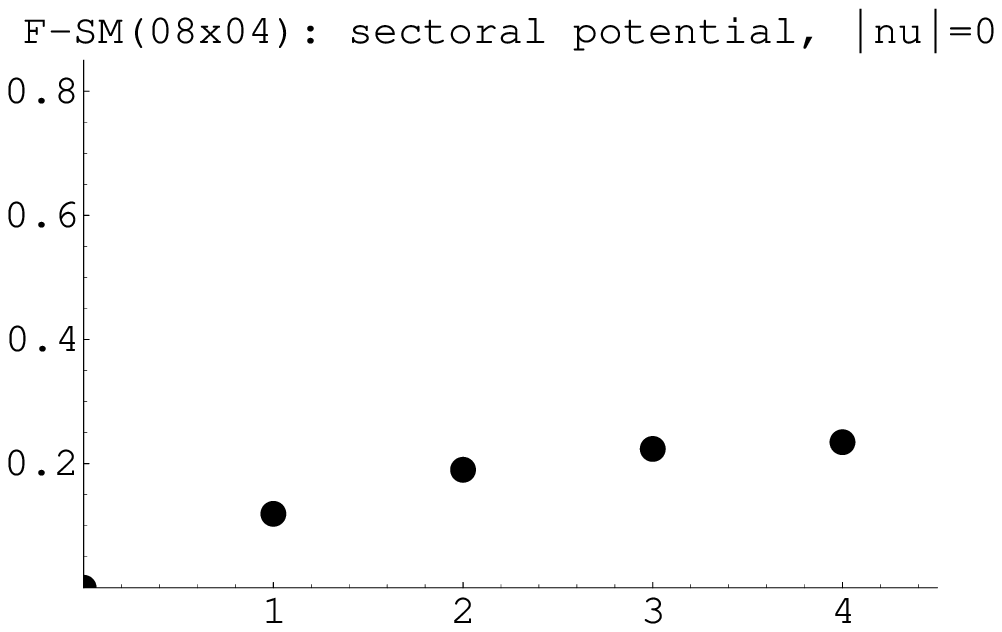,
height=3.4cm,width=4.2cm,angle=90}&
\epsfig{file=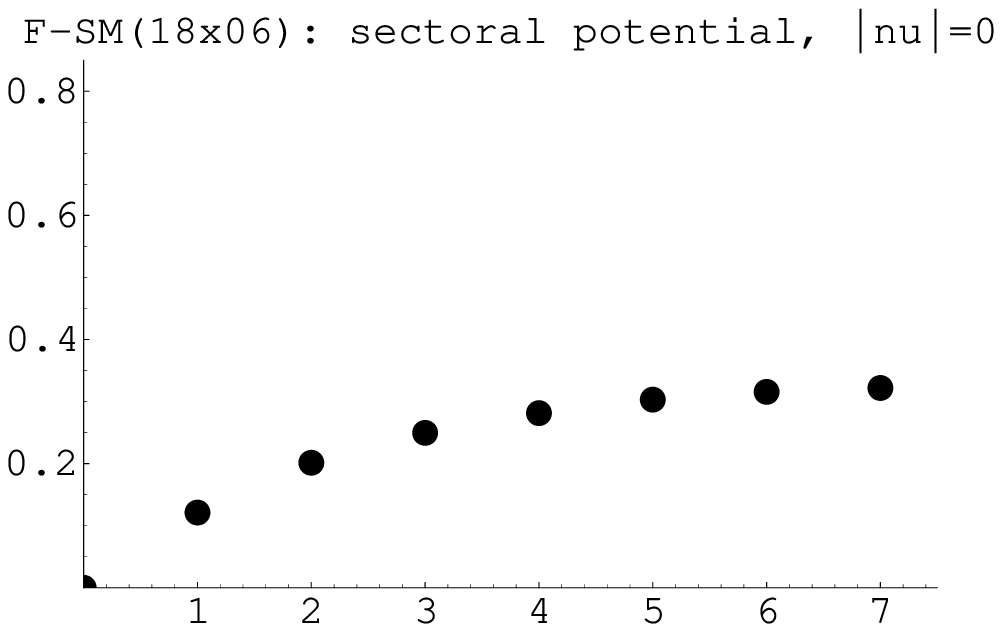,
height=3.4cm,width=4.2cm,angle=90}&
\epsfig{file=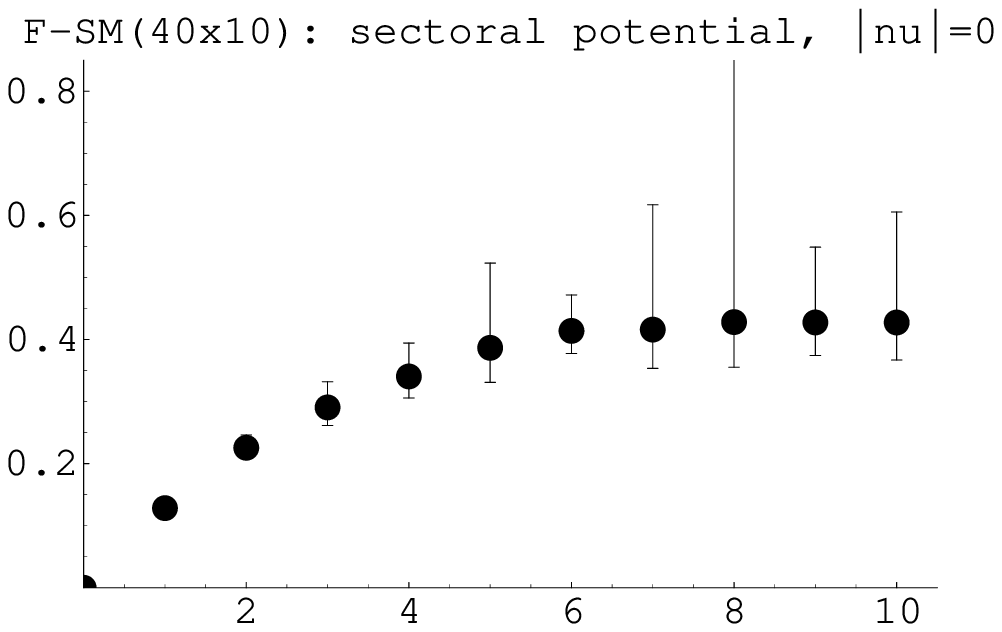,
height=3.4cm,width=4.2cm,angle=90}
\end{tabular}
\vspace*{-2mm}
\caption{\sl\small
Physical heavy quark potential (rightmost column) and sectoral heavy quark
potential on subsets with $|\nu|\!=\!0,1,2,3$ in the small (top), intermediate
(middle) and large (bottom) Leutwyler-Smilga regime.}
\end{figure}

\begin{table}[t]
\begin{center}
\begin{tabular}{|c|ccc|}
\hline
{}&dist=1&dist=2&dist=3\\
\hline
$|\nu|=0$&
$0.119{+0.003\atop -0.003}$&
$0.190{+0.009\atop -0.009}$&
$0.224{+0.013\atop -0.012}$\\
$|\nu|=1$&
$0.222{+0.028\atop -0.025}$&
$0.793{+\infty\atop-0.224}$&
$\mathrm{indet.}$\\
\hline
all conf.&
$0.123{+0.003\atop -0.003}$&
$0.201{+0.010\atop -0.009}$&
$0.240{+0.016\atop -0.015}$\\
\hline
\end{tabular}
\vspace*{-2mm}
\end{center}
\begin{flushleft}
Table~1:~{\sl Sectoral heavy quark potentials and unseparated (physical) poten-
tial at spatial separations $1..3$ in the small Leutwyler Smilga regime
($x\!\ll\!1$).}
\end{flushleft}
\begin{center}
\begin{tabular}{|c|cccc|}
\hline
{}&dist=1&dist=2&dist=3&dist=4\\
\hline
$|\nu|=0$&
$0.121{+0.002\atop -0.002}$&
$0.201{+0.004\atop -0.004}$&
$0.250{+0.006\atop -0.006}$&
$0.281{+0.012\atop -0.011}$\\
$|\nu|=1$&
$0.145{+0.005\atop -0.005}$&
$0.266{+0.015\atop -0.014}$&
$0.388{+0.038\atop -0.031}$&
$0.516{+0.125\atop -0.071}$\\
$|\nu|=2$&
$0.206{+0.030\atop -0.025}$&
$0.658{+\infty\atop-0.305}$&
$\mathrm{indet.}$&
$\mathrm{indet.}$\\
\hline
all conf.&
$0.128{+0.002\atop -0.002}$&
$0.218{+0.008\atop -0.007}$&
$0.278{+0.013\atop -0.012}$&
$0.319{+0.020\atop -0.018}$\\
\hline
\end{tabular}
\vspace*{-2mm}
\end{center}
\begin{flushleft}
Table~2:~{\sl Sectoral heavy quark potentials and unseparated (physical) poten-
tial at separations $1..4$ in the intermediate Leutwyler Smilga regime
($x\!\simeq\!1$).}
\end{flushleft}
\begin{center}
\begin{tabular}{|c|ccccc|}
\hline
{}&dist=1&dist=2&dist=3&dist=4&dist=5\\
\hline
$|\nu|=0$&
$0.128{+0.006\atop -0.006}$&
$0.226{+0.021\atop -0.017}$&
$0.291{+0.041\atop -0.029}$&
$0.340{+0.054\atop -0.035}$&
$0.387{+0.136\atop -0.056}$\\
$|\nu|=1$&
$0.130{+0.004\atop -0.003}$&
$0.229{+0.008\atop -0.007}$&
$0.308{+0.023\atop -0.019}$&
$0.364{+0.028\atop -0.022}$&
$0.407{+0.057\atop -0.036}$\\
$|\nu|=2$&
$0.134{+0.007\atop -0.006}$&
$0.259{+0.032\atop -0.024}$&
$0.383{+0.137\atop -0.056}$&
$0.444{+\infty\atop-0.119}$&
$0.478{+\infty\atop-0.136}$\\
$|\nu|=3$&
$0.145{+0.022\atop -0.018}$&
$0.276{+0.120\atop -0.053}$&
$0.357{+\infty\atop-0.126}$&
$0.457{+\infty\atop-0.223}$&
$\mathrm{indet.}$\\
\hline
all conf.&
$0.130{+0.003\atop -0.003}$&
$0.233{+0.011\atop -0.010}$&
$0.310{+0.018\atop -0.015}$&
$0.366{+0.025\atop -0.020}$&
$0.417{+0.026\atop -0.021}$\\
\hline
\end{tabular}
\vspace*{-2mm}
\end{center}
\begin{flushleft}
Table~3:~{\sl Sectoral heavy quark potentials and unseparated (physical) poten-
tial at spatial separations $1..5$ in the large Leutwyler Smilga regime
($x\!\gg\!1$).}
\end{flushleft}
\end{table}

In the small LS-regime the sectoral potentials for $\nu\!=\!0$ and
$\nu\!=\!\pm1$ differ quite drastically.
In this case the potential in the topologically trivial sector agrees (on a
$1\,\sigma$ basis, cf.\ Tab.\ 1) with the potential on the full sample~--- as
expected, since for $x\!\ll\!1$ the partition function is dominated by $Z_0$.
What might come as a surprise is that the error-bars in the first plot in the
top line of Fig.~5 are {\em smaller\/} than those in the rightmost one, even
though the configurations analyzed in the former case represent a
{\em subset\/} of those used in the latter case.
The reason, as we have seen, is that the complementary subset (the one
comprising the configurations with $\nu\!=\!\pm1$) provides measurement data
which are {\em inconsistent\/} with the data won from the topologically trivial
subset, but has a total weight which is too small to affect the complete
ensemble average significantly (cf.\ the first plot in Fig.~2).
Hence, in the small LS-regime, the tiny contribution from the higher
topological sectors primarily adds some {\em noise\/} to a standard observable
like the heavy quark potential.
The practical recipe to get rid of this effect is to do by hand (for
sufficiently small $x$) what the functional determinant does anyways in the
limit $x\!\to\!0$: Cut away the higher topological sectors !

In the intermediate LS-regime the difference between the two sectoral
potentials for $\nu\!=\!0$ and $\nu\!=\!\pm1$ is smaller than in the previous
regime, but the sectoral potentials are typically several $\sigma$ away
from each other and even the potential in the topologically trivial sector lies
significantly below the unseparated physical potential (cf.\ Tab.\ 2).
From this we see that the sector with $\nu\!=\!0$ does not always yield the
right result.
The intermediate LS-regime shares thus with the small $x$ regime the uneasy
property that different sectors give {\em inconsistent values\/} for physical
observables, but unlike in the small $x$ regime, several sectors yield
{\em sizable contributions\/} to the complete (physical) expectation value~---
hence perfect ergodicity w.r.t.\ $\nu$ is mandatory.

In the large LS-regime differences between expectation values measured on
subsets of definite topological charge diminish further, if the latter differs
only by one (``neighboring sectors'').
On the other hand, in the large $x$ regime there is a multitude of topological
sectors yielding sizable contributions to the physical expectation value, hence
an obvious question is which effect will win in the limit $x\!\to\!\infty$.
Hence, what we would like to know is whether the difference between the heavy
quark potential on the subsets with the lowest and the highest (reasonably
populated) $|\nu|$ will fade away, stay constant or grow if $x$ increases
unlimitedly.
From Tab.\ 3, one gets the impression that the difference between the two
extreme fixed-$|\nu|$ expectation values persists in the large LS-regime.
It seems that the statement by Leutwyler and Smilga that in the large $x$
regime the topological charge has vanishing influence on physical
measurements is reproduced for sufficiently low $|\nu|$ values (in our case
e.g.\ $|\nu|\!\in\!\{0,1,2\}$ but not necessarily for $|\nu|\!=\!3$).
It is also worth noticing that the topologically trivial potential lies at least
$0.5\,\sigma$ below the physical potential.
Hence, a constraint to the topologically trivial sector seems permissible for
$x\!\gg\!1$, though in our simulation a constraint to the sectors $|\nu|\!=\!1$
would have been a better choice.
Of course it is difficult to get definitive answers from numerical data; it may
well be that our LS-value ($x\!\simeq\!4.16$) is still too small for the
large $x$ behaviour to be fully pronounced.
Nevertheless, the situation seems to be analogous to what we have seen in the
case of the heavy quark free energy:
Any differences between expectation values of {\em neighboring sectors\/}
(with sufficiently low $|\nu|$) disappear upon taking the large $x$ limit.
However, even in the large LS-regime no supporting evidence has been found that
$\<V_{q\bar q}\>_\nu$ might be {\em completely independent\/} of $\nu$.


\section{Summary}

The massive multi-flavour Schwinger model has been used to test the statements
by Leutwyler and Smilga about the (ir-)relevance of the topological charge
for QCD in a finite box.
It is expected that an analogous study for full QCD will result in figures
which look, as far as qualitative issues go, exactly the same as the ones
presented here and that, for this reason, the statements below are valid for
QCD too.
The key points are the following:

({\sl i\/})
The three LS-regimes show distinctive properties already on a formal level, by
the way how the contribution per flavour to the total action (the sign-flipped
logarithm of the one-flavour functional determinant) correlates with the
contribution from the gauge field.
Introducing the fermion determinant at fixed $\be$ results in every regime in
an {\em overall suppression\/} of higher topological sectors w.r.t.\ low-lying
ones.
For $x\!\ll\!1$ this suppression is so strong that the functional determinant
effectively acts as a constraint to the topologically trivial sector.
In addition, the sea-flavours effectively result in a {\em sectoral
multiplicative renormalization of\/} $\be$ with a factor greater than 1.
For small and intermediate $x$ this factor is found to decrease as a function
of $|\nu|$, while in the large LS-regime the renormalization factor seems to be
uniform for all topological sectors.

({\sl ii\/})
For the overall distribution of topological charges the predictions by
Leutwyler and Smilga are well observed:
In the regime $x\!\ll\!1$ the partition function (and hence the sample) is
entirely dominated by the topologically trivial sector, i.e.\ the histogram
of topological charges essentially consists of a delta-peak at $\nu\!=\!0$.
In the intermediate regime $(x\!\simeq\!1)$ the distribution is narrow but
nontrivial, i.e.\ sizable contributions stem essentially from the sectors
$\nu\!=\!0,\pm1,\pm2$.
In the regime $x\!\gg\!1$ the topological charge distribution gets broad
and is approximated by a gaussian with width $\<\nu^2\>\simeq x/N_{\!f}$.

({\sl iii\/})
In the small LS-regime ($x\!\ll\!1$) the sectoral expectation value of an
observable in the first topologically nontrivial sector (typically won from
very few configurations) differs drastically from the analogous expectation
value in the topologically trivial sector (represented by the overwhelming
majority of configurations).
This means that a simulation getting stuck at $\nu\!=\!\pm1$ (or even higher)
is absolutely disastrous to the result, i.e.\ tiny non-ergodicities w.r.t.\
the topological charge render an overall sample average meaningless.
The good news is that in this regime the correct distribution is known and
hence there is a simple remedy: A strict constraint to the topologically
trivial sector proves beneficial.

({\sl iv\/})
The intermediate LS-regime ($x\!\simeq\!1$) is characterized
by the distinctive feature that {\em several mutually inconsistent sectors\/}
yield {\em sizable contributions\/} to a standard (i.e.\ not relating to the
$U(1)_A$ issue) physical observable.
This means that the intermediate regime is the one which is most delicate
to simulate in: There is a strong sectoral dependence of physical observables,
yet there is no simple recipe how to guarantee the correct sectoral weighting.
In other words: Physical measurements on a sample won in the intermediate
LS-regime do rely on the simulation algorithm having achieved perfect
ergodicity w.r.t.\ the topological charge.

({\sl v\/})
In the large LS-regime ($x\!\gg\!1$) sectoral averages seem {\em consistent
between neighboring topological sectors\/} (for sufficiently low $|\nu|$), but
no evidence has been found that measurements might be {\em completely
independent\/} of $\nu$.
It seems that jumping (within a well-distributed sample) from the lowest to the
highest accessible $|\nu|$ may affect sectoral averages 
even in the large $x$ regime.
This, if correct, means that the statement by Leutwyler and Smilga according to
which ``the topological charge is an irrelevant concept in the large $x$
regime''~\cite{LeutwylerSmilga} should to be interpreted in the following way:
In the large LS-regime {\em physical observables are immune to modifications
of the $\nu$-histogram which are small compared to its natural width\/}
$(x/N_{\!f})^{1/2}$; they may, however, be sensitive to distortions which go
beyond this limit.
From a lattice perspective the implication is that the result of a full QCD
simulation in the large $x$ regime may be trusted without hesitation as long as
the algorithm did not get stuck in a sector with unreasonably high $|\nu|$, say
at $\nu\!=\!10$ if the simulation was in the regime $x\!\sim\!40$ and
$N_{\!f}\!=\!2$.

({\sl vi\/})
In this work, the transition from the small to the intermediate and the large
LS-regime has been made by adding more sites to the grid while keeping
$\beta, m$ fixed.
It would be interesting to test the {\em universality property\/}, i.e.\
the claim that it is only the product $V\Sigma m$ \cite{LeutwylerSmilga}
which decides on the regime.

({\sl vii\/})
It is worth emphasizing that in the large $x$ simulation used in this work the
predictions by Leutwyler and Smilga turned out to be fulfilled even though the
condition in the r.h.s.\ of (\ref{LScondition}) according to which the pion
would overlap the box was not obeyed, but rather the opposite relation
$1/M_\pi\!\ll\!L$ would hold true, i.e.\ the pion would fit into the box, as is
usual in a lattice context.
Hence the r.h.s.\ of (\ref{LScondition}) seems to be a purely {\em technical
condition\/}, immaterial to the result of the LS-analysis in the large $x$
regime.

({\sl viii\/})
From the fact that the massive multi-flavour Schwinger model follows the
LS-classification even though it resembles QCD slightly above the chiral phase
transition one is led to suspect that this classification might be more
general, even within QCD, than its original derivation indicates.
In other words: The conjecture is that the LS-classification may prove useful
not only in the broken phase but also in the high-temperature phase of QCD~---
as long as one stays in the immediate vicinity of the phase transition,
i.e.\ as long as pions keep being visible (i.e.\ keep dominating the long-range
Green's functions with pion-type quantum numbers; for a recent study see e.g.\
\cite{QCDTARO}).
This, if correct, would imply that the version of the LS-parameter as given in
eqn.\ (\ref{LSPinQCD}) is more general%
\footnote{Note that in the high-temperature phase $\Sigma\!=\!0$, hence
(\ref{LSPdef}) is useless for $T\!>\!T_c$.}
then the original version (\ref{LSPdef}).

({\sl ix\/})
The results of this investigation may be seen as an additional a-posteriori
justification of the resources phenomenological lattice groups have been
provided with in order to be able to study (full) QCD in the large $x$ regime%
\footnote{This is the result of a private attempt to estimate $x$ in dynamical
simulations aiming at the light hadron spectrum from published data (e.g.\
$m_{\rm sea}$ and $a$ in physical units).}.
In addition, they support attempts to develop algorithms which decorrelate the
topological charge faster than standard HMC (see e.g.\ \cite{IlKeMuSt}), as
this is necessary for trustworthy simulations in the intermediate LS-regime,
towards which one moves when comparing different runs with increasing
$\kappa_\mathrm{sea}$ in a fixed physical volume, or even more so, if $\beta$
is kept fixed.
The fact that even in the large $x$ regime full QCD simulations may yield
unreliable results upon generating {\em extreme\/} distortions in the
topological charge distribution should be taken as a motivation to monitor
$\nu$ in all phenomenological studies (i.e.\ not just with an observable
relating to the $U(1)_A$ issue) and, for the time being, as a warning against
pushing the sea-quark mass too light%
\footnote{Here it is assumed that each algorithm has a critical $M_\pi/M_\rh$
(which may well depend on the actions used) beneath which it tends to get stuck
(cf.\ footnote~1).}
(even when increasing the volume together with $\kappa_\mathrm{sea}$ so as to
keep the product $x\!=\!V\Sigma m$ fixed and large compared to~1).
The latter constraint is certainly not a disaster, since there is a
well-elaborate framework \cite{SharpeShoresh} which allows one to extrapolate
from samples generated with a somewhat larger sea-quark mass down to the
physical $M_\pi/M_\rh$.
Hence it seems that our results do represent a little caveat but no real
obstacle to future progress in Lattice QCD.


\subsection*{Acknowledgements}

I would like to thank Steve Sharpe for fruitful physics discussions in an early
phase of this work and Roland Rosenfelder for advice regarding presentation.


\end{document}